\documentclass[lettersize,journal]{IEEEtran}
 \usepackage[table,xcdraw]{xcolor}
\usepackage{amsmath,amsfonts}
\usepackage{algorithm}
\usepackage{algorithmic}
\usepackage{array}
\usepackage{todonotes}
\usepackage{cooltooltips}

\usepackage{textcomp}
\usepackage{stfloats}
\usepackage[export]{adjustbox}
\usepackage{url}
\usepackage{verbatim}
\usepackage{graphicx}
\usepackage[noadjust]{cite}
\usepackage{tikz}
\usepackage{graphicx}
 
\usepackage{amsmath}
\usepackage{arydshln}
\DeclareMathOperator*{\argmax}{\arg\!\max}
\DeclareMathOperator*{\argmin}{\arg\!\min}
\usepackage{amssymb}
\usepackage{amsthm}
\usepackage[hidelinks]{hyperref}
\usepackage[capitalize]{cleveref}
\usepackage{accents}
\usepackage{booktabs}
\usepackage{tabularx}
\usepackage{booktabs}
\usepackage{multirow}
\newcommand{\midsepremove}{\aboverulesep = 0mm \belowrulesep = 0mm}
\midsepremove
\newcommand{\midsepdefault}{\aboverulesep = 0.605mm \belowrulesep = 0.984mm}
\midsepdefault
 \usepackage{graphicx}

\theoremstyle{definition}
\newtheorem{defn}{Definition}
\usepackage{footnote}
\makesavenoteenv{algorithm}
\crefname{definition}{Definition}{Definitions}

\makeatletter
\let\MYcaption\@makecaption
\makeatother
\usepackage[font=footnotesize]{subcaption}
\makeatletter
\let\@makecaption\MYcaption
\makeatother

\usepackage{makecell}
\newcolumntype{x}[1]{>{\centering\arraybackslash}p{#1}}
\usepackage{tikz}
\newcommand\diag[4]{%
  \multicolumn{1}{p{#2}}{\hskip-\tabcolsep
  $\vcenter{\begin{tikzpicture}[baseline=0,anchor=south west,inner sep=#1]
  \path[use as bounding box] (0,0) rectangle (#2+2\tabcolsep,\baselineskip);
  \node[minimum width={#2+2\tabcolsep-\pgflinewidth},
        minimum  height=\baselineskip+\extrarowheight-\pgflinewidth] (box) {};
  \draw[line cap=round] (box.north west) -- (box.south east);
  \node[anchor=south west] at (box.south west) {#3};
  \node[anchor=north east] at (box.north east) {#4};
 \end{tikzpicture}}$\hskip-\tabcolsep}}

\usepackage{bm}
\usepackage{circledsteps}
\newcommand\myCircled[2][]{\ifmmode
\Circled[fill color=black,inner color=white,#1]{#2}
\else
\Circled[fill color=black,inner color=white,#1]{#2}
\fi
}

\makeatletter
\DeclareRobustCommand{\rvdots}{%
  \vbox{
    \baselineskip4\p@\lineskiplimit\z@
    \kern-\p@
    \hbox{.}\hbox{.}\hbox{.}
  }}
\makeatother

\DeclareSymbolFont{largesymbolsA}{U}{txexa}{m}{n}
\DeclareMathSymbol{\varprod}{\mathop}{largesymbolsA}{16}

\DeclareFontFamily{U}{mathx}{\hyphenchar\font45}
\DeclareFontShape{U}{mathx}{m}{n}{
      <5> <6> <7> <8> <9> <10>
      <10.95> <12> <14.4> <17.28> <20.74> <24.88>
      mathx10
      }{}
\DeclareSymbolFont{mathx}{U}{mathx}{m}{n}
\DeclareMathSymbol{\bigtimes}{1}{mathx}{"91}
\usetikzlibrary{calc,angles,quotes,arrows.meta}
\makeatletter
\newcommand{\gettikzxy}[3]{%
	\tikz@scan@one@point\pgfutil@firstofone#1\relax
	\edef#2{\the\pgf@x}%
	\edef#3{\the\pgf@y}%
}
\makeatother
\usepackage{nccmath}
\begin{document}

\title{Tactical Game-theoretic Decision-making with Homotopy Class Constraints}

\author{Michael Khayyat, Alessandro Zanardi, Stefano Arrigoni, Francesco Braghin,~\IEEEmembership{Member,~IEEE}
\thanks{Manuscript received XXXX XX, 2024; revised XXXX XX, 2024. (\textit{Corresponding author: Michael Khayyat.})}

\thanks{Michael Khayyat, Stefano Arrigoni, and Francesco Braghin are with the Department of Mechanical Engineering, Politecnico di Milano, Via La Masa, 1, 20156 Milano MI, Italy. Emails:~\{michael.khayyat\}; \{stefano.arrigoni\}; \{francesco.braghin\} @polimi.it.\\
Alessandro Zanardi is with the Institute for Dynamic Systems and Control, ETH Zurich, 8092 Zurich, Switzerland. Email:~\{azanardi\}@ethz.ch.}}

\newcommand\submittedtext{%
  \footnotesize This work has been submitted to the IEEE for possible publication. Copyright may be transferred without notice, after which this version may no longer be accessible.}

\newcommand\submittednotice{%
\begin{tikzpicture}[remember picture,overlay]
\node[anchor=south,yshift=10pt] at (current page.south) {\fbox{\parbox{\dimexpr0.65\textwidth-\fboxsep-\fboxrule\relax}{\submittedtext}}};
\end{tikzpicture}%
}

\markboth{IEEE \LaTeX Class Journals,~Vol.~xx, No.~xx, XXXXX~2024}%
{XXXXXX \MakeLowercase{\textit{et al.}}: }

\IEEEpubid{0000--0000/00\$00.00~\copyright~2024 IEEE}
\maketitle
\submittednotice


\begin{abstract}
We propose a tactical homotopy-aware decision-making framework for game-theoretic motion planning in urban environments. We model urban driving as a generalized Nash equilibrium problem and employ a mixed-integer approach to tame the combinatorial aspect of motion planning. More specifically, by utilizing homotopy classes, we partition the high-dimensional solution space into finite, well-defined subregions. Each subregion (homotopy) corresponds to a high-level tactical decision, such as the passing order between pairs of players. The proposed formulation allows to find global optimal Nash equilibria in a computationally tractable manner by solving a mixed-integer quadratic program. Each homotopy decision is represented by a binary variable that activates different sets of linear collision avoidance constraints. This extra homotopic constraint allows to find solutions in a more efficient way (on a roundabout scenario on average 5-times faster). We experimentally validate the proposed approach on scenarios taken from the \textit{rounD} dataset. Simulation-based testing in receding horizon fashion demonstrates the capability of the framework in achieving globally optimal solutions while yielding a 78\% average decrease in the computational time with respect to an implementation without the homotopic constraints.
\end{abstract}


\section{Introduction}
\label{sec:introduction}
As autonomous vehicles navigate through dynamic environments, the need for intelligent decision-making becomes paramount. Traditional motion planning approaches, which employ predict-\textit{then}-plan pipelines, do not take into account the mutual dependence of all vehicles’ decisions. They often struggle in environments characterized by complex interactions among road agents, resulting in an overly conservative or passive behavior, also known as the \textit{frozen robot problem}~\cite{trautman2010unfreezing}. In interaction-rich environments, the coupling between vehicles' decisions poses a significant challenge to motion planning. Some approaches attempt to remedy this problem by improving the quality of predictions by harnessing the power of deep-learned motion forecasting models~\cite{liebner2012driver,salzmann2020trajectron, bansal2018chauffeurnet, yuan2021agentformer, pmlr-v100-chai20a}. Alternative techniques utilize forward simulation models, where a transition model of the environment describes how the traffic scene evolves due to actions taken by the agent in a probabilistic manner~\cite{9526613,8569729,7795596}. Other methodologies exploit game-theoretic models to generate interactive behavior by solving a joint trajectory prediction and planning problem~\cite{le2022algames,zanardi2021urban,zhu2022sequential,evens2022learning,fridovich2020efficient,britzelmeier2019numerical,laine2023computation,dreves2019best}. This work falls into the latter category, where we leverage the principles of game theory and homotopy planning~\cite{bhattacharya_2021,chen2023interactive} to generate globally optimal collision-free motion plans in urban environments.

\begin{figure}[H] 
 \includegraphics[width=\linewidth]{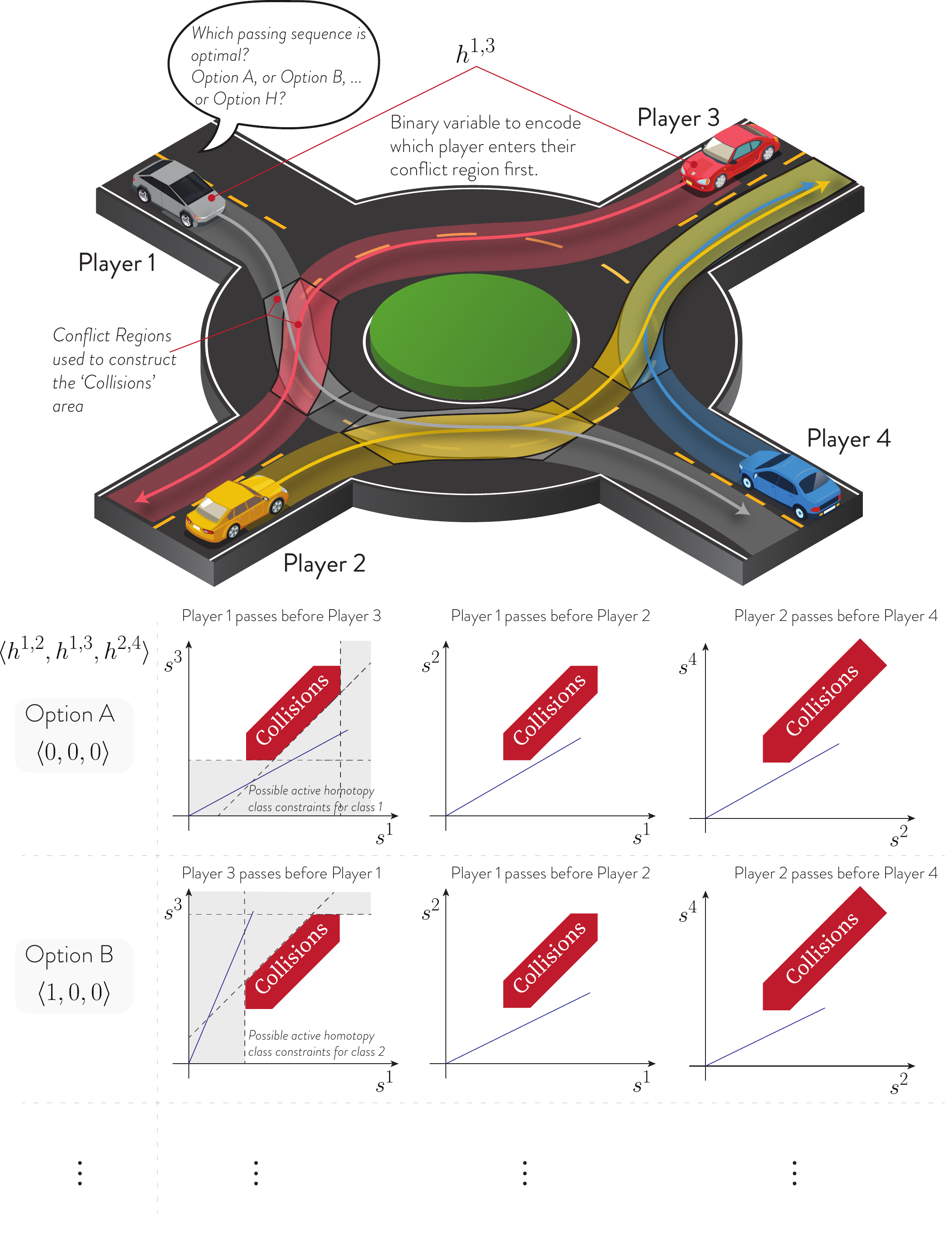}   
    \caption{An example of a roundabout scenario with four players. The non-convex problem is represented in Frenet coordinate frame (lower half) which allow to easily incorporate homotopy class constraints to determine the optimal passing sequences. 
    The conflict regions are used to construct collision avoidance areas and the binary variables $h^{1,2},h^{1,3}$, and $h^{2,4}$ encode the order in which vehicles enter their conflict regions, representing different homotopy classes. 
    The lower section showcases two different tactical options and their corresponding trajectories. In the $s^1-s^3$ progress plane, fixing an homotopy class corresponds to activate different sets of linear constraints.}
    \label{fig:scenarioWithHomtopyIndication}
\end{figure}
\IEEEpubidadjcol

This work lies in the domain of non-cooperative game theory, where agents are self-interested, yet not adversarial, with some shared objectives - e.g., nobody wants to collide. We investigate urban driving scenarios where vehicles follow pre-planned reference paths, such as that depicted in \cref{fig:scenarioWithHomtopyIndication}, while adjusting their speed to account for interactions with other vehicles. The multiagent trajectory planning problem is formulated as a generalized Nash equilibrium problem (GNEP)~\cite{facchinei2010generalized} with linear shared collisions constraints encoded using a mixed-integer quadratic programming (MIQP) framework. Building on the cooperative structure of urban driving~\cite{10460127,evens2022learning, zanardi2021urban}, we recast the problem as a generalized mixed-integer potential game (GMIPG)~\cite{facchinei2011decomposition,sagratella2017algorithms}, a specialized form of generalized potential games (GPGs) which are a subclass of GNEPs with desirable convergence properties. To manage the computational complexity of this problem, we introduce homotopy class constraints to tame the combinatorial aspect. More specifically, by leveraging homotopy principles, we partition the solution space into distinct subregions, each representing a sequence of high-level tactical decisions, such as the order in which vehicles enter their conflict regions, as demonstrated in \cref{fig:scenarioWithHomtopyIndication}. This formulation allows us to transform the trajectory planning problem into a more manageable form, facilitating an exhaustive yet computationally efficient search over all possible passing orders to identify the globally optimal solution. Finally, the MIQP is integrated within a model predictive control (MPC) framework and tested in a roundabout scenario extracted from the \textit{rounD} dataset~\cite{rounDdataset} to validate its performance and global optimality guarantees.

\subsection{Related Work}

Our work relates to several topics in the game-theoretic planning and prediction literature.

\subsubsection{Game-theoretic Motion Planning}
In the realm of dynamic non-cooperative game theory, road agents are viewed not merely as passive entities navigating predefined paths, but as active participants in a dynamic game of strategy~\cite{bacsar1998dynamic}. There exist several solution concepts for predicting the outcome of the strategic interactions of a game. These include minimax regret, correlated equilibrium, and trembling-hand perfect equilibrium, among others. However, the most widely adopted solution concept is the Nash equilibrium (NE)~\cite{leyton2022essentials,fischer2014generalized}. At the NE, no agent can improve its outcome by unilateral deviations from its strategy, and thus being a self-enforcing strategy, it can be quite attractive to seek.

Game-theoretic approaches have proven to be effective for modeling decision-making and have been successfully applied in various applications, namely urban driving~\cite{evens2022learning,zanardi2021urban,fridovich2020efficient,9341137,britzelmeier2019numerical, fisac2019hierarchical}, and racing~\cite{williams2016aggressive,spica2020real,wang2021game,liniger2019noncooperative,notomista2020enhancing}. A substantial body of work employs iterated best response algorithms to find a Nash equilibrium in the joint trajectory space of all agents\cite{williams2016aggressive,spica2020real,wang2021game, 9992957, 10178143,dreves2019best,fabiani2019multi}, and nonlinear programming (NLP) techniques dedicated to solving the  coupled optimal control problems (OCPs) of GNEPs~\cite{zhu2022sequential,le2022algames,singh2023globally,fridovich2020efficient}. Other approaches exploit the intrinsic cooperative nature of urban driving to give the trajectory planning problem additional structure. This allows reducing the OCPs to a single one within a GPG~\cite{monderer1996potential} framework~\cite{10460127, la2016potential,zanardi2021urban,evens2022learning}, and solving for a pure-strategy Nash equilibrium using a suitable off-the-shelf solver. This work aligns with the latter approach, as we aim for a more refined notion of equilibria~\cite{monderer1996potential,carbonell2014refinements} - one that is optimal from the social point of view as it introduces some notions of fairness~\cite{la2016potential}. In a decentralized multi-agent setting, the solution obtained by solving the GPG is likely to be found by the players, thus it enhances strategy alignment \cite{multiplenash} in the game.

In the above-mentioned related work, the employed shared collision constraints are generally non-convex. This presents a computational complexity in solving the GNEP and presents challenges regarding the optimality of the solution. Several attempts to tackle this issue are present in the literature. For example, in~\cite{britzelmeier2021decomposition}, a penalty approach is utilized to absorb the collision constraints in the cost function, which simplifies computational complexity but comes with the drawback that the penalty parameter must go to infinity to ensure collision avoidance. In~\cite{katriniok2017distributed}, a semidefinite programming (SDP) approach is employed then a hierarchy based on fixed rules is computed; however, this approach results in a rank constraint that is dropped in the relaxed problem. Alternatively, convexification techniques have been employed in~\cite{8263811,liu2018convex, zhang2022receding, schulman2014motion, schulman2013finding}. While such techniques attempt to remedy the problem, they bring forth several challenges: the resulting approximations may compromise collision avoidance guarantees, with only local optimality being ensured~\cite{le2022algames,mazumdar2019finding,wang2021game,zhu2022sequential,evens2022learning}. Additionally, meticulous handcrafting of the initial guess is required, given the inherent sensitivity of NLP solvers to it, as highlighted in~\cite[Section 5.6]{le2022algames}.

Mixed-Integer programming (MIP) allows integrating discrete decision-making within continuous optimization frameworks \cite{quirynen2023real}. Additionally, it offers a powerful framework for computing optimal trajectories with linear hard collision avoidance constraints~\cite{schouwenaars2002safe, 7795555, schouwenaars2001mixed}. Advanced software solutions such as Gurobi\textsuperscript{\texttrademark}~\cite{gurobi} have played a pivotal role in making MIP a viable choice for medium-term trajectory planning and high-level tactical decision-making \cite{schouwenaars2002safe, quirynen2023tailored}. MIP has been successfully employed in multi-agent systems. For instance, in \cite{7535369,10.1145/3566097.3567849} propose a distributed MIP approach that efficiently handles the scheduling of autonomous vehicles at (multiple) intersections. Similarly, \cite{9667332} extends the application of MILP to include trajectory smoothing, thus guaranteeing that autonomous vehicles not only navigate intersections efficiently, but also improve passenger comfort and reduce fuel consumption. In \cite{10178275}, Caregnato-Neto et al. propose a novel line-of-sight constraint in MIP models to ensure connectivity between agents in a multi-agent systems, enhancing the tractability and efficiency of MIP for motion planning algorithms. In \cite{10156567}, Bhattacharyya et al. propose an optimal control and inverse optimal control based method for interactive motion planning in lane merging scenarios. The non-convex nature of the feasible regions and lane discipline is handled by introducing integer decision variables, thereby resulting in a MIQP. Furthermore, MIP has been successfully integrated in various game-theoretic motion planning frameworks. For instance, in~\cite{kessler2020linear, fabiani2019multi}, a multi-agent planning problem is formulated as a differential game and solved within an MIQP framework, where decisions are optimized to ensure efficient and safe trajectories by minimizing a joint cost function that accounts for individual agent goals and players joint dynamics. In~\cite{8814060}, a tree-based approach is employed, and the optimal trajectory is solved within a mixed-integer linear programming (MILP) framework. While in~\cite{7995795}, a hierarchical MIQP scheme is employed for a multi-agent planning problem. We refer the reader to~\cite{ioan2021mixed} for a comprehensive survey of MIP motion planning approaches for both single-agent and multi-agent settings.


\subsubsection{Homotopy Planning}

Motion planning is intrinsically an NP-hard problem~\cite{hopcroft1984complexity}. This complexity arises from the vast space of possible control actions, which requires decomposition techniques to achieve practical solutions within feasible time frames~\cite{schouwenaars2002safe}. In the domain of game-theoretic motion planning, several attempts have been made to improve the computational tractability of the resulting game by employing divide-and-conquer approaches. For example,~\cite{10384147,9981416} introduces factorization techniques in which the game is decomposed into multiple independent subgames. Other works decompose the motion planning problem using topology-based clustering approaches, such as path homotopy~\cite{bhattacharya_2021, bhattacharya_2011}, as in~\cite{7182335, 7535369}, where the homotopy class is defined as a set of trajectories that share start and end points (relaxed in~\cite{chen2023interactive}) which can be continuously deformed into each other without intersecting an obstacle. Additionally, path-velocity decomposition (PVD)~\cite{kant1986toward} is adopted to address the complexities of multi-agent motion planning~\cite{gregoire2014priority}. In PVD, a collision-free driving corridor is obtained~\cite{8574950,8917274,8814241,7182335,schafer2021computation}, and then the vehicle velocity is optimized along it~\cite{7225909}. In this work, we integrate the concept of homotopy class in a PVD setting. More specifically, assuming prior knowledge of the paths vehicles travel on, we employ the notion of homotopy to divide the joint planning space into distinct classes. This allows us to systematically address interactions between vehicles and plan optimal joint trajectories accordingly.

\subsection{Contributions}

In this work, we develop a game-theoretic motion planning framework for autonomous vehicles in urban environments. We address the prevalent lack of global optimality guarantees \cite{fridovich2020efficient,le2022algames,gupta2023game} of game-theoretic frameworks by proposing a homotopy-based tactical planner that solves the Nash equilibria of urban dynamic games.

In particular, we consider a setting in which a predefined path is associated with each player. This setup allows us to implement linear collision avoidance constraints in the Frenet reference frame using a mixed-integer methodology. Furthermore, we embed the notion of homotopy in the formulation to encode high-level tactical decisions between pairs of players, effectively linking these strategies to the collision avoidance constraints. This approach enables us to partition the planning space into well-defined subregions, thereby facilitating an exhaustive search over qualitatively different equilibria of the game, such as different passing sequences. Although the introduction of homotopy classes into our framework requires additional binary variables, one for each pair of conflicting paths, these variables establish dependencies among previously independent binary variables that govern collision avoidance. This interdependence enormously facilitates the branch-and-bound search. Finally, we observe that the exhaustive search over all possible tactical decisions facilitates a distributed implementation that does not suffer from the equilibrium selection problem. Indeed, different agents are less likely to independently find different local equilibria -- up to practically rare hypersymmetric cases. 

Experimentally, we show that in various common urban scenarios, such as roundabouts and lane merging, our solver can effectively identify all socially optimal Nash equilibria trajectories of the GMIPG by iterating over all possible homotopy classes. Moreover, we show that our framework can solve for the optimal homotopy class alongside its Nash equilibrium trajectory within a computational time comparable to that required for iterating over a fixed homotopy class. This efficiency is facilitated by the structured partitioning of the solution space, which optimizes the computational process. In addition, we show that tactical reasoning about homotopy classes and their resulting coherency constraints reduces the computational time by $78\%$ on average in our scenarios.

\subsection{Paper Organization}
The remainder of this paper is organized as follows.  \cref{subsec:backgroundandproblemformulation} provides a thorough overview of GNEPs and GPGs alongside an outline of the general problem formulation. In \cref{subsec:homotopy}, the concept of homotopy classes is introduced in the problem formulation, which lays the foundation for the introduction of the collision constraints formulation. Subsequently, in \cref{subsec:modellingofagents}, the modelling of vehicles, along with the associated objective function is introduced. In \cref{sec:simulationresults}, the performance of the MIQP problem formulation is assessed by testing it in a roundabout scenario. Finally, \cref{sec:conclusion} presents remarks and outlines promising future research directions.

\section{Preliminaries}
\label{subsec:backgroundandproblemformulation}

In a dynamic game consisting of a set of $A$ rational players, labeled by the index set $\mathcal{A} = \{1,2,\dots,A \}$, every player $\nu \in \mathcal{A}$ makes decisions on their control inputs $u^\nu_{k} \ \forall k \in [0,\dots,N-1]$ over a planning horizon of length $N$. The state vector of each player at time-step $k$, $x^\nu_{k}$, evolves according to the following dynamics:

\begin{equation}
\label{eq:dynamicsinit}
    x^\nu_{k+1} = f^\nu(x_k^\nu, u_k^\nu), \ \forall k \in [0,\dots,N-1].
\end{equation}

We define the augmented state and control vectors for player $\nu$ as $x^\nu := [x^\nu_k]_{k=0}^{N}$ and $u^\nu := [u^\nu_k]_{k=0}^{N-1} $, respectively. We denote the state vector of all players at time step $k$ as $x_k$, the control input vector at time step $k$ as $u_k$, and define the augmented state and control vectors of all players as $x := [x_k]_{k=0}^{N}$ and $u := [u_k]_{k=0}^{N-1} $. For the sake of notational simplicity, we adopt the notation $\square ^{-\nu}$ from~\cite{facchinei2010generalized} to refer to the quantity $\square$ corresponding to players defined by the set $\mathcal{A}\backslash\{\nu\}$, that is, comprising all players except $\nu$. 

Additionally, we denote the decision variables for player $\nu$ by $z^\nu := (x^\nu, u^\nu, q^\nu)$, where $q^\nu$ are any additional variables that arise due to the problem formulation, such as the homotopy variables in our work. Thus, we can define the following constraints:

\begin{equation}
    z^\nu \in \mathcal{Z}^\nu (z^{-\nu}),
\end{equation}

such that
\begin{equation}
\begin{split}
    \mathcal{Z}^\nu(z^{-\nu}) = \{ z^\nu=(x^\nu, u^\nu, q^\nu) \ | & \ g^\nu_{\text{i}}(z^\nu, z^{-\nu}) \leq 0,  \\
    & \ g^\nu_{\text{e}} (z^\nu, z^{-\nu}) = 0  \},
\end{split}
\end{equation}
where $g^\nu_{\text{i}}(z^\nu)$ collects all the inequality constraints of player $\nu$, such as the path constraints which include collision avoidance and state and actuation limits, and $g^\nu_{\text{e}} (z^\nu)$ encodes the equality constraints such as those that govern the dynamics, as established by (\ref{eq:dynamicsinit}).

Each player attempts to find a strategy $ z^\nu \in \mathcal{Z}^\nu (z^{-\nu})$ that minimizes some time-additive cost function $J^\nu (z^\nu, z^{-\nu})$. To this end, the trajectory optimization problem, for agent $\nu$ is is expressed in the following compact form:
\IEEEpubidadjcol 
    \begin{equation}
    \label{eq:gnep}
P^\nu \left\{
\begin{split}
&\underset{\substack{z^\nu \in \mathcal{Z}^\nu(z^{-\nu})}}{\min} \ J^\nu (z^\nu, z^{-\nu}) & \\[2ex]
&\text{s.t }  \ x_{0} = x_{\text{init}}
\end{split}
\right.
\end{equation}
where $x_{\text{init}}$ is the initial state of the players.

Let $S^\nu(z^{-\nu})$ denote the set of solutions of $P^\nu$. Then, the following definition can be formulated as follows~\cite{britzelmeier2019numerical}:
\begin{defn}(\textit{Generalized Nash Equilibrium Problem (GNEP)}) \\
    Find $\hat{z}$ with $\hat{z}^\nu \in S^\nu (\hat{z}^{-\nu})  \ \forall \nu \in \mathcal{A}$.
\end{defn}


As highlighted in \cref{sec:introduction}, the solution concept we seek for the dynamic game is the (pure-strategy) Nash equilibrium strategy, denoted as $ z^\star \in \mathcal{Z}$.

\begin{defn}
    (\textit{Pure-Strategy Nash Equilibrium}) The pure-strategy profile $z^{\star} = (z^{1,\star},\dots,z^{M,\star})$ is a pure strategy Nash equilibrium if and only if $\forall \nu \in \mathcal{A}$ the following holds:
    \begin{equation}
    \begin{split}
    &J^\nu(z^\star) = J^\nu(z^{\nu,\star},z^{-\nu,\star})  \leq  J^\nu(\bar{z}^{\nu}, z^{-\nu,\star}), \\
    &\quad\quad \forall  \bar{z}^{\nu} = \mathcal{Z}(z^{-\nu,\star}).
    \end{split}
    \end{equation}
In other words, no player can improve its outcome by unilaterally perturbing its strategy in a locally feasible direction.
\end{defn}

Computing the NE of (\ref{eq:gnep}) is non-trivial, as it is a joint optimization problem for all the players. To this end, we seek to exploit the structure of the cost function $J^\nu$ and thus give the game some additional structure. In a more specific context, we aim to leverage the cooperative structure inherent in urban driving, wherein the personal cost function $J^\nu$ depends on the states and controls of player $\nu$ only\footnote{It is noteworthy that such a characteristic may not find relevance in racing applications, where the strategic objective of player $\nu$ extends beyond merely minimizing their own personal costs to simultaneously maximizing the costs incurred by other players.}, i.e $J^\nu(z^\nu)$.

Generalized potential games (GPGs) are a subclass of GNEPs~\cite{sagratella2017algorithms}. GPGs are characterized by possessing a pure-strategy Nash equilibrium, with the players (unknowingly) optimizing the same potential function~\cite{facchinei2011decomposition, shapley1953stochastic}.   

\begin{defn}(\textit{Generalized Potential Game})
\label{def:wpg}
A GNEP is a generalized potential game if and only if there exists a potential function $F(z): \mathcal{Z} \mapsto \mathbb{R}$, such that $\forall \nu \in \mathcal{A}$:
\begin{equation*}
\begin{split}
    & J^\nu(z^{\nu}, z^{-\nu})-  J^\nu(\hat{z}^{\nu}, z^{-\nu}) = F(z^{\nu}, z^{-\nu})-  F(\hat{z}^{\nu}, z^{-\nu}), \\
    & \quad \quad \forall z^{\nu}, \hat{z}^{\nu} \in \mathcal{Z}^\nu, z^{-\nu} \in \mathcal{Z}^{-\nu}.
\end{split}
\end{equation*}
Thus, the Nash equilibria sets of a game where every player $\nu$ is minimizing $J^\nu$ is equivalent to that where every player is minimizing $F$\footnote{For proof, readers are referred to~\cite[Chapter 2.2]{la2016potential}}. 
\end{defn}


\section{Homotopy-based Game Theoretic Motion Planning}
In this section, we present our methodology and problem formulation. We begin by defining path envelopes and conflict regions in \cref{subsec:pecr}, which lay the foundation to define the notion of homotopy classes in \cref{subsec:homotopy} and introduce the constraints formulation, which encompasses collision constraints, inter-sample collision constraints, precedence constraints, dynamics constraints, control constraints, and finally present the overall problem formulation.

\subsection{Path Envelopes and Conflict Regions}
\label{subsec:pecr}
To effectively manage vehicle interactions in typical urban environments, such as intersections and lane merges, we introduce a framework that allows us to systematically define potential areas of interaction between vehicles. Our approach, which builds on the topological relationships between vehicle paths presented in~\cite{zhan2017spatially}, is underpinned by the introduction of the concepts of ``path envelopes" and ``conflict regions" for vehicles.

\begin{figure}[H]
    \centering
    \begin{subfigure}{0.33\linewidth}
    \includegraphics[scale=0.5]{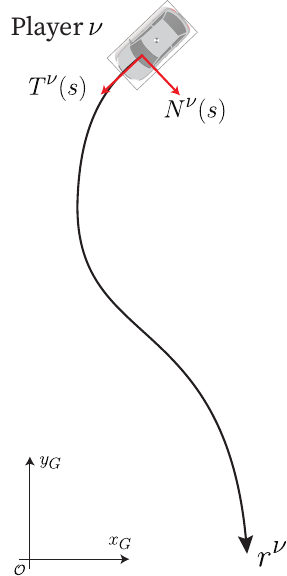}
    \caption{}
    \label{fig:envelope1}
    \end{subfigure}%
    \begin{subfigure}{0.33\linewidth}
    \includegraphics[scale=0.5]{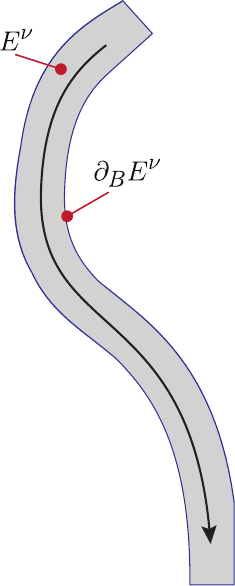}
    \caption{}
    \label{fig:envelope2}
    \end{subfigure}%
    \begin{subfigure}{0.33\linewidth}
    \includegraphics[scale=0.5]{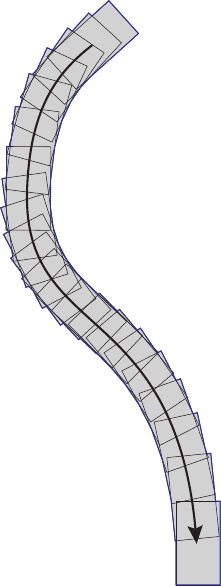}
    \caption{}
    \label{fig:envelope3}
    \end{subfigure}
    \caption[Path Envelope Definition]{(a) Shows a player $\nu$ traversing a path $r^\nu(s)$ defined by vectors $T^\nu(s)$ and $N^\nu(s)$. (b) Shows the path envelope $E^\nu$ and its boundary $\partial_B E^\nu$. (c) Shows an approximation of the envelope and boundary as a result of the discretization procedure. }
\end{figure}

The path envelope represents the area that encapsulates the entire trajectory of the vehicle as it traverses a path. Mathematically, it is defined as follows.

\begin{defn}{(\textit{Path Envelope})}\label[definition]{defpath} The path envelope for player $\nu$, denoted as $E^\nu$, is defined as the union of all half-spaces, $\mathcal{H}^\nu(s)$, that characterize the vehicle as it traverses a path $r^\nu (s)$, as depicted in \cref{fig:envelope1,fig:envelope2}. The path is characterized by the unit normal vector $N^\nu(s)$, and the unit tangential vector $T^\nu(s)$, which uniquely defines the orientation of the vehicle, $\psi^\nu$, in the global reference frame, $\mathcal{O}$. Thus, the envelope can be defined as follows:

\begin{equation*}
\begin{split}
    E^\nu = \bigcup_{s= 0}^{l_s^\nu} \mathcal{H}^\nu(s),
    \end{split}
\end{equation*}

such that the occupancy of the vehicle with length $l^\nu$ and width $w^\nu$ on the road is represented using the half-space ($H$-representation) as follows:

\begin{equation}
\label{eq:collisonavoidance}
    \mathcal{H}^\nu(s) = \{ x \in \mathbb{R}^2~|~A^\nu(s) x \leq b^\nu(s) \},
\end{equation}
where
\begin{align*}
A^\nu(s) &= 
\begin{bmatrix}
\sin(\psi^\nu(s)) & -\cos (\psi^\nu(s)) \\
-\sin(\psi^\nu(s)) & \cos (\psi^\nu(s)) \\
\cos(\psi^\nu(s)) & \sin (\psi^\nu(s)) \\
-\cos(\psi^\nu(s)) & -\sin (\psi^\nu(s)) \\
\end{bmatrix},\\
b^\nu(s) &= \begin{bmatrix}
    w^\nu /2\\
    w^\nu/2 \\
    l^\nu/2 \\
    l^\nu/2
\end{bmatrix} + A^\nu r^\nu (s).
\end{align*}

\end{defn}

The path envelope boundary is the outer edge of the path envelope, as depicted in \cref{fig:envelope2}. It is accurately defined as follows:

\begin{defn}{(\textit{Path Envelope Boundary})}\label[definition]{defpathbound}  The boundary for player $\nu$, denoted as $\partial E^\nu$, for an envelope $E^\nu \subseteq \mathbb{R}^2$ is defined as the intersection of the closure of $E^\nu$ with the closure of its complement \cite{mendelson1990introduction}:

\begin{equation*}
    \partial_B E^\nu := \overline{S} \cap \overline{(\mathbb{R}^2 \backslash E^\nu)}.
\end{equation*}

\end{defn}

A conflict region is an area where the path envelopes of two players overlap, indicating a segment along each player's path where collisions could occur with the other. The conflict region for two players, $\nu$ and $\mu$, can be systematically defined as follows.

\begin{figure}[H]
    \centering
    \begin{subfigure}{0.25\textwidth}
    \includegraphics[width = \linewidth]{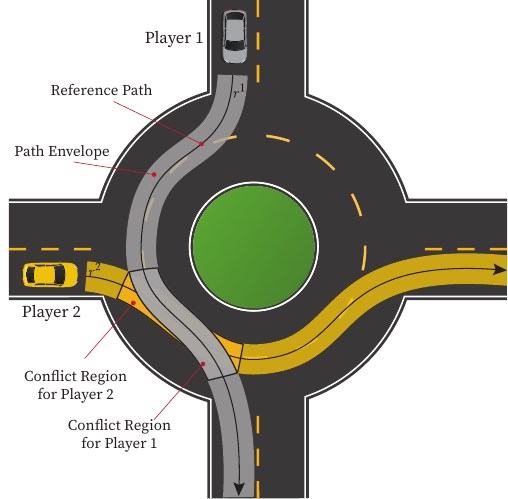}
    \caption{}
    \label{fig:path_envelope}
    \end{subfigure}%
    \begin{subfigure}{0.25\textwidth}
    \includegraphics[width = \linewidth]{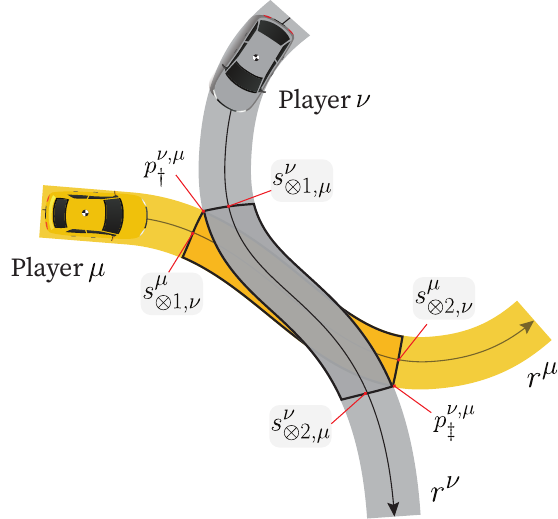}
    \caption{}
    \label{fig:conflictregion}
    \end{subfigure}
    \caption[Conflict region definition]{(a) Shows the conflict region for two players navigating an intersection, (b) A close-up of the conflict regions highlighting relevant points for its definition}
\end{figure}

\begin{defn}{(\textit{Conflict Region})} \label[definition]{defconflicT}  A conflict region between two players $\nu$ and $\mu$ travelling along paths $r^\nu(s)$ and $r^\mu(t)$ respectively exists if $\partial_B E^\nu \cap \partial_B E^\mu \neq \emptyset$. The conflict region for player $\nu$ is characterized by two points $p^{\nu,\mu}_\dagger$ and $p^{\nu,\mu}_\ddagger$, which mark its beginning and end, as illustrated in \cref{fig:conflictregion}, and are expressed as follows:

\begin{equation*}
\begin{split}
    p^{\nu,\mu}_\dagger = \argmin_p s  \text{ such that } p \in \partial_B E^\nu \cap \partial_B E^\mu, \\
    p^{\nu,\mu}_\ddagger = \argmax_p s \text{ such that } p \in \partial_B E^\nu \cap \partial_B E^\mu. \\    
\end{split}
\end{equation*} 

The projection of these points on $r^\nu$, obtained by minimizing the norm of the distance between these points to $r^\nu$, are denoted as $s^\nu_{\otimes1,\mu}$, and $s^\nu_{\otimes2,\mu}$. Then, the conflict region for player $\nu$ due to its interaction with player $\mu$ can be effectively defined as 

\begin{equation*}
\begin{split}
    E^{\nu,\mu}_\text{cr} = \{(x,y) \in E^\nu ~ | ~ s^\nu_{\otimes1,\mu} \leq \Pi_{r^\nu} (x,y) \leq s^\nu_{\otimes2,\mu} \},
    \end{split}
\end{equation*}
where $\Pi_{r^\nu} (x,y)$ is the projection operator for a point $(x,y)$ onto the parametrized path $r^\nu$.
\end{defn}

\subsection{Homotopy Classes and Constraints Formulation}
\label{subsec:homotopy}
In the context of urban driving where vehicles are following predefined reference paths, the principal challenge in high-level tactical decision-making lies in determining the precedence among vehicles. This process involves determining which vehicle should proceed before the other at intersections or merge points, i.e the passing sequence. Consequently, the motion planning problem can be decomposed into two steps: enumerating the possible passing sequences within a scenario and calculating the optimal trajectory within each sequence. 

To systematically define the passing sequence of vehicles, we employ the concept of homotopy classes from~\cite{bhattacharya_2021}, and adapt it with a flexible interpretation to generalize it to dynamic environments. More specifically, we relax the requirement of defining homotopies according to joint start and end points and resort instead to the sequence of entry into conflict regions.
\begin{defn}{(\textit{Homotopy Class})}
\label{defn:homotopy}
Two joint motion plans of an ordered pair of players $[\nu,\mu]$ belong to the same homotopy class if the sequence of the players' entries into their respective conflict regions is identical.
\end{defn}
To this end, we define the binary parameter $h^{\nu,\mu}$ to represent the homotopy class, as follows:
\begin{equation}
    h^{\nu,\mu} = \begin{cases}
        0 & \text{if player $\nu$ enters first}, \\
        1 & \text{if player $\mu$ enters first}. 
    \end{cases}
\end{equation}

Note that only one binary homotopy class variable is introduced for the ordered pair of players $[\nu,\mu]$ to prevent the redundancy of defining both $h^{\nu,\mu}$ and $ h^{\mu,\nu}$. Otherwise, we would need to impose additional equality constraints, specifically $h^{\nu,\mu} = 1 - h^{\mu,\nu}$, to ensure that each homotopy class is uniquely defined.

\subsubsection{Collision Avoidance Constraints}
\label{subsub:collisons}
\begin{figure}[H]
\begin{subfigure}{0.25\textwidth}
    \centering
    \includegraphics[width=\linewidth]{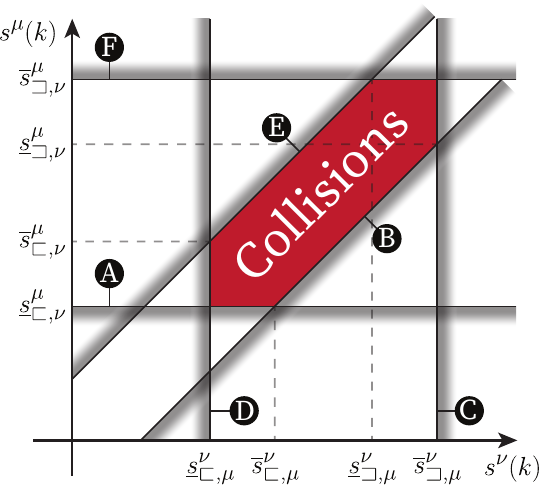}   
    \caption{}
    \label{fig:homotopy_diagram_a}
\end{subfigure}%
\begin{subfigure}{0.25\textwidth}
    \centering
    \includegraphics[width=\linewidth]{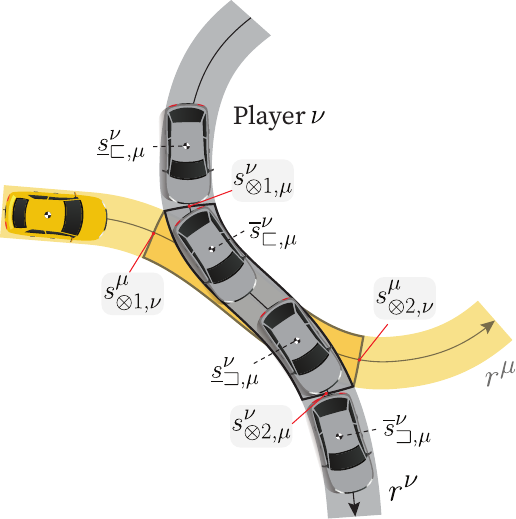}   
    \caption{}
    \label{fig:homotopy_diagram_b}
\end{subfigure}
  \caption{(a) Defines the area where an $(s^\nu(k), s^\mu(k))$ pair is a collision for the players $\nu$ and $\mu$ in the $s^\nu, s^\mu$ plane. (b) Shows the corresponding limits of the collisions area of the homotopy diagram in the global world frame.}
  \label{fig:homotopy_diagram} 
\end{figure}

The collisions area in \cref{fig:homotopy_diagram_a} represents $(s^\nu(k), s^\mu(k))$ pairs which lead to a collision for players  $\nu$ and $\mu$, and it is geometrically defined using the results obtained from \cref{subsec:pecr}.  In fact, once $s^{\nu}_{\otimes1,\mu}, s^\nu_{\otimes2,\mu},
s^\mu_{\otimes1,\nu},$ and $s^\mu_{\otimes2,\nu}$ are obtained as described in \cref{defconflicT}, the bounds of the collision area can be computed as follows for the case where $\text{sgn}(s^\nu_{\otimes2,\mu} - s^\nu_{\otimes1,\mu}) = \text{sgn}(s^\mu_{\otimes2,\nu} - s^\mu_{\otimes1,\nu})$ \footnote{This yields the most general case of the constraints. The other case where $\text{sgn}(s^\nu_{\otimes2,\mu} - s^\nu_{\otimes1,\mu}) \neq \text{sgn}(s^\mu_{\otimes2,\nu} - s^\mu_{\otimes1,\nu})$ is touched upon in \cref{subsub:specialcases}}:
\begin{equation}
\label{eq:bounds}
\begin{aligned}
        \overline{s}^{\nu}_{\sqsubset,\mu} &= s^\nu_{\otimes1,\mu} - \tfrac{1}{2}\text{len}^\nu, \\
        \underline{s}^{\nu}_{\sqsubset,\mu} &= s^\nu_{\otimes1,\mu} + \tfrac{1}{2}\text{len}^\nu,\\
        \overline{s}^{\nu}_{\sqsupset,\mu} &= s^\nu_{\otimes2,\mu} - \tfrac{1}{2}\text{len}^\nu,\\
        \underline{s}^{\nu}_{\sqsupset,\mu} &= s^\nu_{\otimes2,\mu} + \tfrac{1}{2}\text{len}^\nu,\\
\end{aligned}
\qquad
\begin{aligned}
        \overline{s}^{\mu}_{\sqsubset,\nu} &= s^\mu_{\otimes1,\nu} - \tfrac{1}{2}\text{len}^\mu,\\
        \underline{s}^{\mu}_{\sqsubset,\nu} &= s^\mu_{\otimes1,\nu} + \tfrac{1}{2}\text{len}^\mu,\\
        \overline{s}^{\mu}_{\sqsupset,\nu} &= s^\mu_{\otimes2,\nu} - \tfrac{1}{2}\text{len}^\mu,\\
        \underline{s}^{\mu}_{\sqsupset,\nu} &= s^\mu_{\otimes2,\nu} + \tfrac{1}{2}\text{len}^\mu.\\
\end{aligned}
\end{equation}

\begin{figure*}[t]
\begin{subfigure}{0.5\textwidth}
    \centering
    \includegraphics[width=\linewidth]{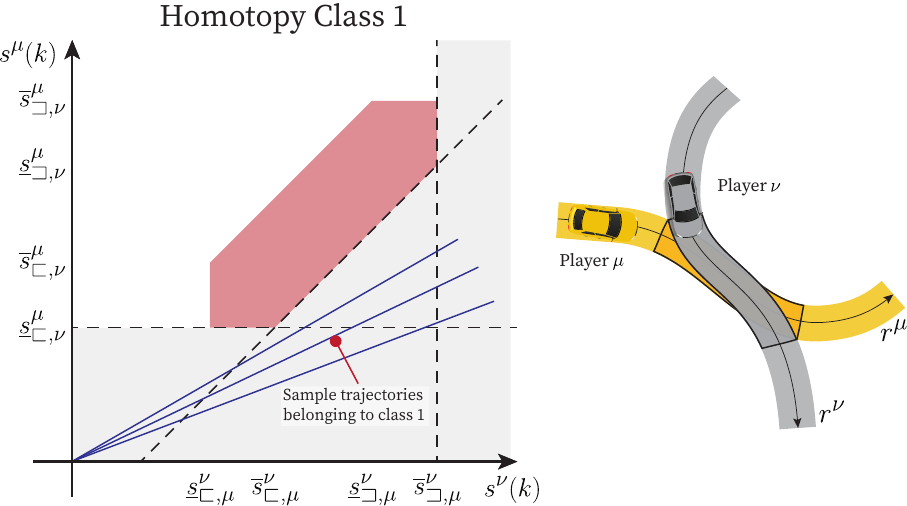}   
    \caption{}
    \label{fig:homotopy_2classes_a}
\end{subfigure}%
\begin{subfigure}{0.5\textwidth}
    \centering
    \includegraphics[width=\linewidth]{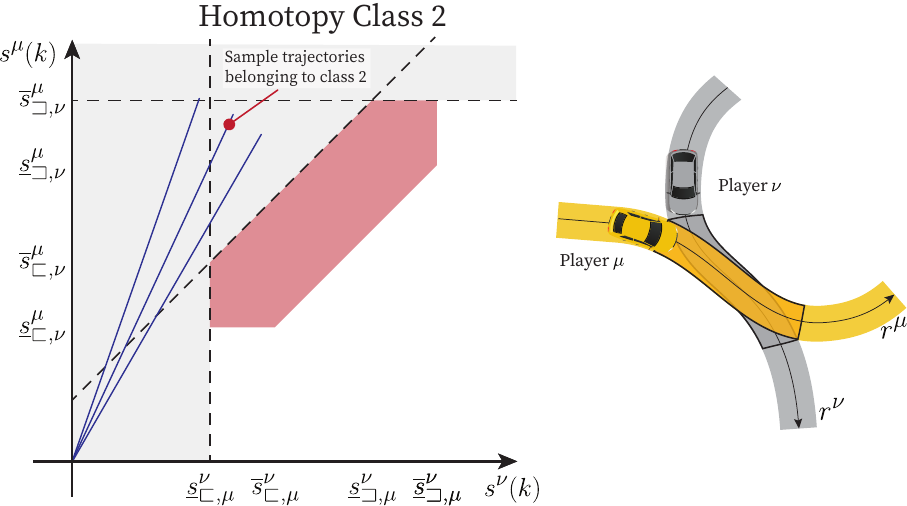}   
    \caption{}
    \label{fig:homotopy_2classes_b}
\end{subfigure}
\caption{(a), (b) Show trajectories that belong to homotopy class 1 and 2 respectively, highlighting the active constraints in each case and the passing sequence.}
\label{fig:homotopy_2classes}
\end{figure*}

Consequently, it is possible to establish the collision avoidance constraints by ensuring that any $(s^\nu(k), s^\mu(k))$ pair lies outside the designated collision area, thus adhering to at least one of the labeled inequality constraints detailed in \cref{fig:homotopy_diagram_a}. Collision avoidance between \textit{combinations} of any two players $\nu, \mu$ whose path envelopes intersect is formulated as a Mixed-Integer Problem (MIP). We introduce six binary variables, denoted as $\sigma^{\nu,\mu}_A, \sigma^{\nu,\mu}_B, \sigma^{\nu,\mu}_C, \sigma^{\nu,\mu}_D, \sigma^{\nu,\mu}_E$, and $\sigma^{\nu,\mu}_F$, where each variable corresponds to a specific inequality constraint, i.e \myCircled{A} - \myCircled{F}, respectively. The value of each binary variable indicates whether the associated inequality constraint is active. It should be noted that the integration of homotopy classes with collision avoidance is a fundamental aspect of our formulation. Specifically, for homotopy class 1, the pair $(s^\nu(k), s^\mu(k))$ must satisfy at least one of the inequality constraints \myCircled{A}, \myCircled{B}, or \myCircled{C}, and for homotopy class 2, it must satisfy at least one of \myCircled{D}, \myCircled{E}, or \myCircled{F}. This concept is illustrated in~\cref{fig:homotopy_2classes}, where the trajectory of a vehicle along its reference path is plotted against that of the other. As demonstrated in \cref{fig:homotopy_2classes_b}, all joint trajectories that fall below the designated \textit{collisions} area are classified into one homotopy class, while those above it belong to another homotopy class. Refer to \cref{fig:additionalhomotopy} for additional illustrations on the boundaries of the collision region. Thus, the logical statement of the homotopy class can be converted to the following:

\begin{equation}
    \begin{split}
     \sigma^{\nu,\mu}_A (k) + \sigma^{\nu,\mu}_B (k) + \sigma^{\nu,\mu}_C (k)    &=  (1 - h ^{\nu,\mu}), \\
   \sigma^{\nu,\mu}_D (k) + \sigma^{\nu,\mu}_E(k) + \sigma^{\nu,\mu}_F(k)      &= h ^{\nu,\mu}.
    \end{split}
    \label{eq:logical}
\end{equation}

Utilizing the Big-$M$ formulation~\cite{nocedal1999numerical}, the collision-avoidance constraints can be formulated as follows between two players $\nu$ and $\mu$:

\begin{subequations}
\begin{align}
s^\nu (k) &\leq \underline{s}^{\nu}_{\sqsubset,\mu} + M\cdot(1-\sigma^{\nu,\mu}_D(k)), \label{eqn:collision_a} \\
s^\mu (k) &\leq \underline{s}^{\mu}_{\sqsubset,\nu} + M\cdot(1- \sigma^{\nu,\mu}_A(k)), \label{eqn:collision_c} \\
s^\nu (k) &\geq \overline{s}^{\nu}_{\sqsupset,\mu} - M\cdot(1-\sigma^{\nu,\mu}_C(k)), \label{eqn:collision_e} \\
s^\mu (k) &\geq \overline{s}^{\mu}_{\sqsupset,\nu} - M\cdot(1-\sigma^{\nu,\mu}_F(k)), \label{eqn:collision_g} \\
s^\nu (k) &\leq s^{\mu}(k) - (\overline{s}^{\mu}_{\sqsubset,\nu} - \underline{s}^{\nu}_{\sqsubset,\mu} ) + M\cdot(1-\sigma^{\nu,\mu}_E(k)), \label{eqn:collision_i}\\
s^\mu (k) &\leq s^{\nu}(k) + (\underline{s}^{\mu}_{\sqsubset,\nu} - \overline{s}^{\nu}_{\sqsubset,\mu} ) + M\cdot(1-\sigma^{\nu,\mu}_B(k)). \label{eqn:collision_k}
\end{align}
\label{eq:collision}
\end{subequations}

where $M$ is a constant with a sufficiently large value. To further reduce the search space, additional constraints are added. As specified in (\ref{eq:stateandcontrolconstraints}), the progress is guaranteed to be non-decreasing for all vehicles. Thus, it is possible to translate the monotonicity of the progress to that of binary variables $\sigma^{\nu,\mu}_A$, $\sigma^{\nu,\mu}_C$, $\sigma^{\nu,\mu}_F$, and $\sigma^{\nu,\mu}_D$ by introducing the following interstage inequalities: 
\begin{equation}
    \begin{split}
        \sigma^{\nu,\mu}_A (k+1) - \sigma^{\nu,\mu}_A (k) \leq 0, \\
        \sigma^{\nu,\mu}_C (k+1) - \sigma^{\nu,\mu}_C (k) \geq 0, \\
        \sigma^{\nu,\mu}_F (k+1) - \sigma^{\nu,\mu}_F (k) \geq 0, \\
        \sigma^{\nu,\mu}_D (k+1) - \sigma^{\nu,\mu}_D (k) \leq 0. \\
    \end{split}
    \label{eq:monotonicity}
\end{equation}

Note that $\sigma^{\nu,\mu}_B$ and $\sigma^{\nu,\mu}_E$ are not necessarily monotonic on all the planning horizon $N$. However, the impact on the search space is minimal, primarily due to the monotonicity of the other binary variables and the inequality constraints imposed by (\ref{eq:logical}). This formulation results in a Mixed-Integer Quadratic Problem (MIQP). Note that an alternative formulation of collision avoidance is presented in \cref{appendix:scenario_b}.

\subsubsection{Inter-sample Collision Avoidance} Due to the introduced discretization, collision avoidance is ensured at every time step $k \in [1,\dots,N]$. However, this raises a critical consideration regarding the selection of the discretization time-step $dT$. Given a fixed prediction horizon time, i.e $dT\cdot N$, a choice of a large $dT$ would result in a small $N$ thereby resulting in a fast computational time, but potentially infeasible solutions will be generated. Conversely, a small $dT$ would enhance inter-sample constraint enforcement, but could significantly increase computational time. Several methodologies exist to better ensure inter-sample collision avoidance. The first technique involves augmenting the size of the collisions area. An alternative technique, based on~\cite{richards2015inter} involves the introduction of additional constraints. More specifically, inter-sample avoidance can be realized by directly applying identical avoidance constraints from time step $k$ at the preceding time step $k-1$. This ensures that whichever constraint is selected at time $k$ is also enforced at time $k-1$, as follows:

\begin{subequations}
\begin{align}
s^\nu (k-1) &\leq \underline{s}^{\nu}_{\sqsubset,\mu} + M\cdot(1-\sigma^{\nu,\mu}_D(k)) \label{eqn:collision_a}, \\
s^\mu (k-1) &\leq \underline{s}^{\mu}_{\sqsubset,\nu} + M\cdot(1- \sigma^{\nu,\mu}_A(k)) \label{eqn:collision_c}, \\
s^\nu (k-1) &\geq \overline{s}^{\nu}_{\sqsupset,\mu} - M\cdot(1-\sigma^{\nu,\mu}_C(k)) \label{eqn:collision_e}, \\
s^\mu (k-1) &\geq \overline{s}^{\mu}_{\sqsupset,\nu} - M\cdot(1-\sigma^{\nu,\mu}_F(k)) \label{eqn:collision_g}, \\
s^\nu (k-1) &\leq s^{\mu}(k-1) - (\overline{s}^{\mu}_{\sqsubset,\nu} - \underline{s}^{\nu}_{\sqsubset,\mu} ) + M\cdot(1-\sigma^{\nu,\mu}_E(k)) \label{eqn:collision_i},\\
s^\mu (k-1) &\leq s^{\nu}(k-1) + (\underline{s}^{\mu}_{\sqsubset,\nu} - \overline{s}^{\nu}_{\sqsubset,\mu} ) + M\cdot(1-\sigma^{\nu,\mu}_B(k)) \label{eqn:collision_k}.
\end{align}
\label{eq:collision2}
\end{subequations}

Note that with respect to (\ref{eq:collision}), in (\ref{eq:collision2}) the index on the states changed from $k-1$; however, that of the binary variables remained $k$. This ensure that no point on the line connecting $(s^\nu(k-1), s^\mu(k-1))$ and $(s^\nu(k), s^\mu(k))$ can be inside the collisions area.

\subsubsection{Special Cases}
\label{subsub:specialcases}
\begin{figure}
\centering
\begin{subfigure}{0.5\textwidth}
    \centering
    \includegraphics[width=\linewidth]{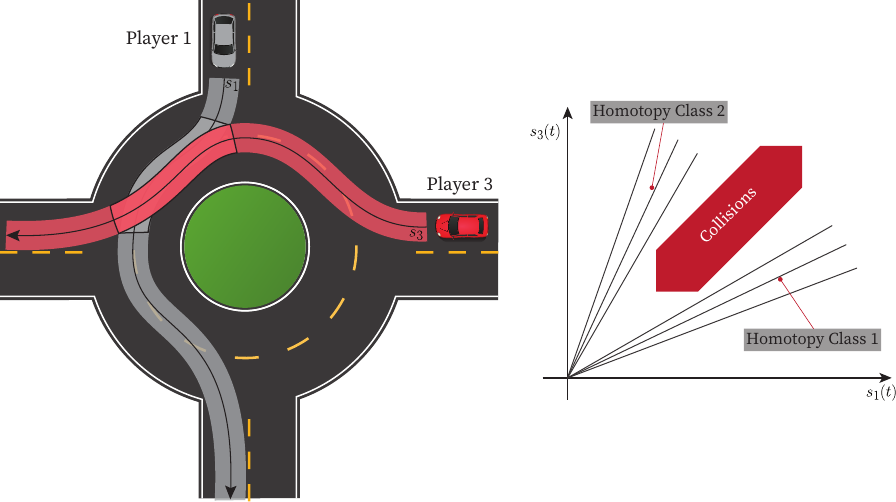}   
    \caption{}
    \label{fig:homotopy_otherscenarios_a}
\end{subfigure}
\begin{subfigure}{0.5\textwidth}
    \centering
    \includegraphics[width=\linewidth]{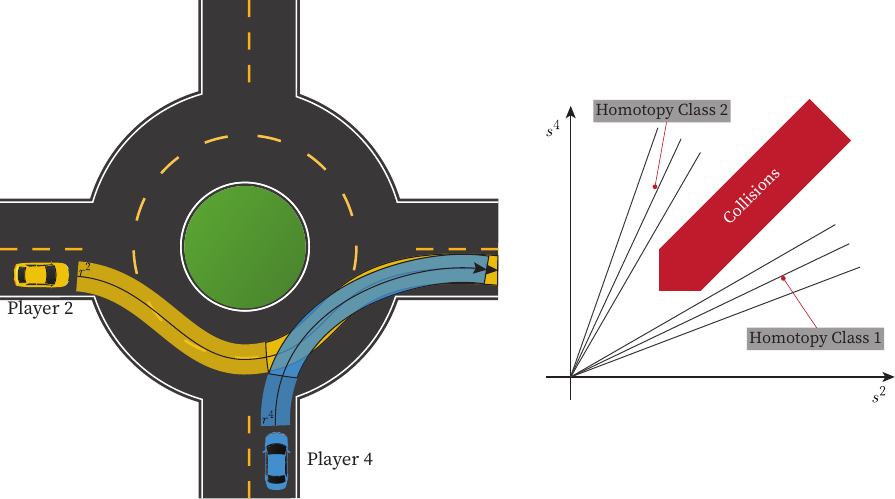}   
    \caption{}
    \label{fig:homotopy_otherscenarios_b}
\end{subfigure}
\begin{subfigure}{0.5\textwidth}
    \centering
    \includegraphics[width=\linewidth]{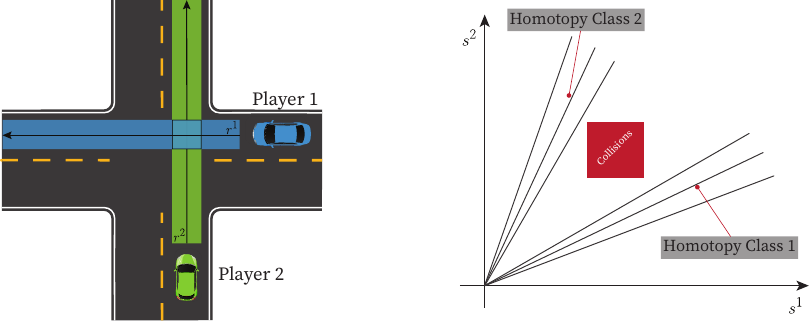}   
    \caption{}
    \label{fig:homotopy_otherscenarios_c}
\end{subfigure}
\begin{subfigure}{0.5\textwidth}
    \centering
    \includegraphics[width=\linewidth]{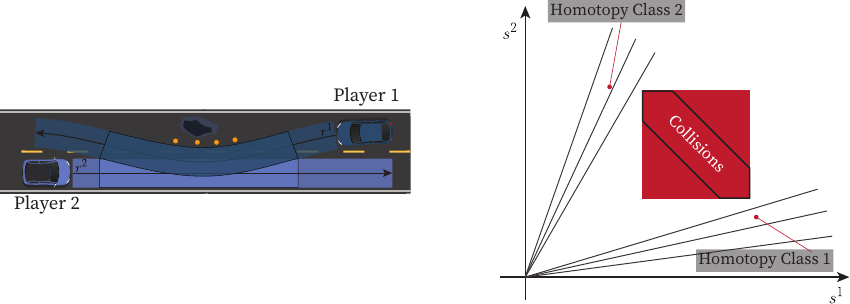}   
    \caption{}
    \label{fig:homotopy_otherscenarios_d}
\end{subfigure}
\caption{Showcases the homotopy diagram and the collisions area for various cases. (a) Shows the most general case. (b) Shows the case where the players merge into the same lane. (c) Shows the case where the players paths intersect at a single point. (d) Shows the case where players are travelling in opposite directions. }
\label{fig:homotopy_otherscenarios} 
\end{figure}

This scenario shown in \cref{fig:homotopy_2classes} represents the most general case of the collision avoidance constraints. Other cases, including those that arise from interactions in straight-line intersections, lane changes, and lane merges, can be considered as specific instances of it. Various scenarios along their collision avoidance regions in the $s^\nu, s^\mu$ plane are presented in \cref{fig:homotopy_otherscenarios}. While \cref{fig:homotopy_otherscenarios_a} represents an identical formulation of the collision avoidance constraints, \cref{fig:homotopy_otherscenarios_b} represents the case where \myCircled{C} and \myCircled{F} are dropped. Additionally, \cref{fig:homotopy_otherscenarios_c} represents the case where $\underline{s}^{\nu}_{\sqsubset,\mu} = \underline{s}^{\nu}_{\sqsupset,\mu} $, $\overline{s}^{\nu}_{\sqsupset,\mu} = \overline{s}^{\nu}_{\sqsubset,\mu} $,$\overline{s}^{\mu}_{\sqsupset,\nu} = \overline{s}^{\nu}_{\sqsubset,\nu} $, and $\overline{s}^{\mu}_{\sqsupset,\nu} = \overline{s}^{\mu}_{\sqsubset,\nu} $, thus \myCircled{B} and \myCircled{E} are dropped.

\cref{fig:homotopy_otherscenarios_d} illustrates a scenario where $\text{sgn}(s^\nu_{\otimes2,\mu} - s^\nu_{\otimes1,\mu}) \neq \text{sgn}(s^\mu_{\otimes2,\nu} - s^\mu_{\otimes1,\nu})$. The collisions area can be decomposed into two parts. First, the area enclosed by the black boundary represents the geometric collisions area, which essentially mirrors the configuration depicted in \cref{fig:homotopy_otherscenarios_a}. Second, the effective region, shown in red, emerges as a consequence of the constraints outlined in (\ref{eq:stateandcontrolconstraints_b}). For further clarification, refer to \cref{appendix:scenario_d}.

\subsubsection{Extension to N-Vehicle Scenarios}

\begin{figure}[htb!]
\begin{subfigure}{0.25\textwidth}
    \centering
    \includegraphics[width=\linewidth]{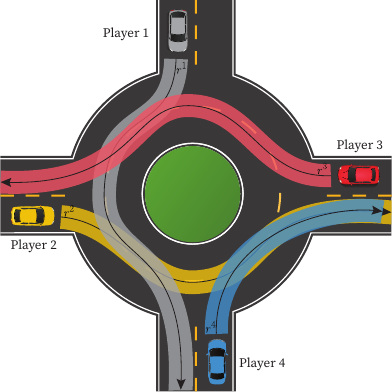}   
    \caption{}
    \label{fig:homotopy_scenario1}
\end{subfigure}%
\begin{subfigure}{0.25\textwidth}
    \centering
    \includegraphics[scale=0.8]{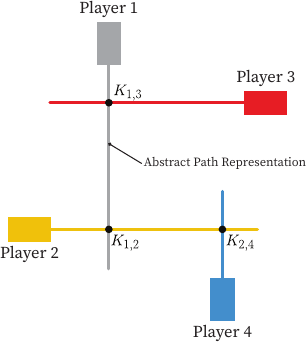}   
    \caption{}
    \label{fig:homotopy_scenario2}
\end{subfigure}
    \caption[Real Scenario vs Abstract Representation]{(a) Real scenario (b) Abstract representation that is used to iterate the possible homotopy classes}
  \label{fig:homotopy_scenario_defn} 
\end{figure}

To extend the definition of homotopy class to cases with more than two vehicles, we consider the example illustrated in \cref{fig:homotopy_scenario1}. The abstract representation of this scenario is shown in \cref{fig:homotopy_scenario2}, where each knot $K_{\nu,\mu}$ indicates an intersection between the path envelopes of players $\nu$ and $\mu$. By applying \cref{defn:homotopy} to every pair of players whose reference paths result in a knot, namely $\big\{ \{1,2\}, \{1,3\}, \{2,4\} \big\}$, the scenario homotopy class can be defined as follows:

\begin{defn}{(\textit{Scenario Homotopy Class})}
A scenario homotopy class is defined as an ordered tuple, $\langle [h^{\nu,\mu}]_{\nu \in \mathcal{A}, \mu \in \mathcal{I}^\nu} \rangle$, where each element $h^{\nu,\mu}$ corresponds to the homotopy class of an ordered pair of agents $[\nu,\mu]$. 
\end{defn}
For the remainder of the manuscript, we denote by $\mathcal{I}^\nu \subset \mathcal{A}$ the set of players whose path envelope intersects with that of player $\nu$.

Thus, for the scenario depicted in \cref{fig:homotopy_scenario_defn}, the scenario homotopy class is represented as $\langle h^{1,2}, h^{1,3}, h^{2,4} \rangle$. Note that in the case of $n$ knots, the number of all possible scenario homotopy classes is $2^n$.

\subsubsection{Precedence Constraints}
\begin{figure}
    \centering
    \includegraphics[width=0.8\linewidth]{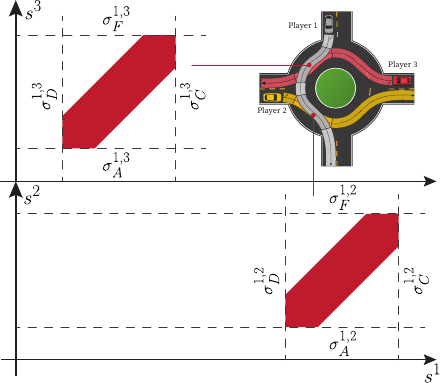}
    \caption{Shows how the precedence constraints are constructed for the interactions of player 1 in the highlighted roundabout scenario.}
    \label{fig:precedence}
\end{figure}

Note that in the case where the path of player $\nu$ intersects with two or more other players,  the binary variables $\sigma^{\nu,\bullet}_\square$ are not independent, and the relationship between them depends purely on the scenario-specific arrangement of the paths followed by the vehicles. This realization allows us to introduce additional constraints to the problem to promote branch pruning in the solver. We demonstrate this fact for Player 1 in the scenario illustrated in \cref{fig:homotopy_scenario1}. The collision area in the $(s^1, s^2)$ and $(s^1, s^3)$ planes is plotted in \cref{fig:precedence}.

Note that the following conditions could be inferred:

\begin{subequations} \label{eqn:precedence}
    \begin{align}
        &\sigma^{1,3}_D(k) = 1 \ \wedge  h^{1,2} = 0 \ \implies \sigma^{1,2}_A(k) = 1 \label{eqn:precedence_a},\\
        &\sigma^{1,3}_D(k) = 1 \ \wedge  h^{1,2} = 1 \ \implies \sigma^{1,2}_F(k) + \sigma^{1,2}_D(k) = 1  \label{eqn:precedence_b}, \\
        &\sigma^{1,2}_C(k) = 1 \ \wedge h^{1,3} = 0 \ \implies \sigma^{1,3}_C(k) + \sigma^{1,3}_A(k) = 1 \label{eqn:precedence_c}, \\
        &\sigma^{1,2}_C(k) = 1 \ \wedge h^{1,3} = 1 \ \implies \sigma^{1,3}_F(k) = 1. \label{eqn:precedence_d}        
    \end{align}
\end{subequations}

Conditions (\ref{eqn:precedence_a}) and (\ref{eqn:precedence_b}) dictate that if $s_1 \leq  \underline{s}^{1}_{\sqsubset,3}$, then this implies that $s_1 \leq  \underline{s}^{1}_{\sqsubset,2}$. By checking which constraints should be active in this region when  $s_1 \leq  \underline{s}^{1}_{\sqsubset,2}$, then it could be seen that If $h^{1,2} =0$, then this means that $\sigma^{1,2}_A(k) = 1$. Otherwise, $\sigma^{1,2}_F(k) + \sigma^{1,2}_D(k) = 1$.

These conditions can be expressed using the big-$M$ formulation as follows for (\ref{eqn:precedence_a}) and (\ref{eqn:precedence_b}):

\begin{subequations} \label{eqn:precedence_math}
    \begin{align}
        & \sigma^{1,2}_A(k) \leq 1 + M\cdot h^{1,2} + M\cdot(1-\sigma^{1,3}_D(k)) \label{eqn:precedence_math_a},\\
        & \sigma^{1,2}_A(k) \geq 1- M\cdot h^{1,2} - M\cdot(1-\sigma^{1,3}_D(k)) \label{eqn:precedence_math_b},\\    
        & \sigma^{1,2}_F(k) + \sigma^{1,2}_D(k) \leq 1 +  M\cdot(1- h^{1,2}) + M\cdot(1-\sigma^{1,3}_D(k)) \label{eqn:precedence_math_c},\\   
        & \sigma^{1,2}_F(k) + \sigma^{1,2}_D(k) \geq 1 -  M\cdot(1- h^{1,2}) - M\cdot(1-\sigma^{1,3}_D(k)). \label{eqn:precedence_math_c}      
    \end{align}
\end{subequations}



\subsection{Modelling of Agents}
\label{subsec:modellingofagents}
\subsubsection{Vehicle Dynamics}
\label{subsec:dynamics}
In scenarios where vehicles follow predefined reference paths, leveraging the Frenet coordinate system is advantageous for several reasons~\cite{britzelmeier2021decomposition}. This coordinate system facilitates the parameterization of the vehicle's trajectory using clothoids~\cite{bertolazzi2015g1, silva2018clothoid}, polynomials~\cite{werling2010optimal}, or splines~\cite{mercy2017spline}. Furthermore, the decoupling of longitudinal and lateral dynamics simplifies the design of control and planning algorithms~\cite{werling2010optimal, 9951715}. 

It is important to note that, throughout our analysis, we assume that the vehicle follows the path precisely without any lateral deviation. While this assumption is limiting, it yields several favorable properties regarding the system model and the structure of the nonlinear programming problem (NLP). Moreover, it aligns with motion planning standards, where deviations from the centerline of reference paths are typically penalized. Consequently, the dynamics of the vehicles can be effectively represented using a linear point-mass model\footnote{For additional vehicle models that utilize the Frenet coordinate system, the reader is referred to~\cite{kloeser2020nmpc, reiter2021parameterization,9341731,novi2019real,chen2020mpc,7810277,esterle2020optimal}.}, which captures the progress of the center of gravity (CoG) along the reference path, indicated by the parameter $s$. It should be emphasized that the pose of the vehicle can be uniquely characterized by $s$. The discrete linear dynamics of vehicle $\nu$ is expressed as follows:

\begin{equation}
\label{eq:dynamics}
    x^\nu(k+1) = \begin{bmatrix}
                  1 & dT \\
                  0 & 1
              \end{bmatrix}  x^\nu(k) + \begin{bmatrix}
                  0 \\ dT
              \end{bmatrix} u^\nu(k),
\end{equation}

where $  x^\nu(k) = \begin{bmatrix} s^\nu(k) \\ \dot s^\nu (k) \end{bmatrix} $ and $u^\nu(k) = \begin{bmatrix}
    \ddot{s}^\nu(k)
\end{bmatrix}.$

\subsubsection{State And Control Constraints}

The feasible domains for states and control inputs at time step $k$ for player $\nu$ are defined as follows:

\begin{subequations}
\label{eq:stateandcontrolconstraints}
    \begin{align}
        0 \leq &~s^\nu(k) \leq \infty \label{eq:stateandcontrolconstraints_a},\\
        0 \leq &~\dot{s}^\nu(k) \leq \overline{\text{vel}^\nu}_{\text{long}} \label{eq:stateandcontrolconstraints_b},\\
        \underline{\text{acc}^\nu}_{\text{long}} \leq~&\ddot{s}^\nu(k) \leq \overline{\text{acc}^\nu}_{\text{long}} \label{eq:stateandcontrolconstraints_c}.
    \end{align}
\end{subequations}

\subsubsection{Personal Cost Function}

We assume that all players want to make progress as quickly as possible while reducing control effort. Thus, we define the following personal cost functions for every player $\nu \in \mathcal{A}$

\begin{equation}
\label{eq:costfunc}
    J^\nu(z^\nu) = \sum_{k=0}^{N-1} u^\nu(k)^\intercal P^\nu u^\nu(k) - r^\nu(s^\nu(N) - s^\nu(0)),
\end{equation}

where $P^\nu$ is a positive semidefinite matrix and $r^\nu$ is a non-negative scalar.

\subsection{Resulting Mixed-Integer Quadratic Program}
Now that we have established all its necessary components, the generalized Nash equilibrium problem (GNEP) that governs the dynamic game can be formulated.

\subsubsection{GNEP Formulation}
Following the notation adopted for the state and control constraints, we introduce $h^\nu$ as the concatenation of all homotopy variables associated with player $\nu$, and $\sigma^\nu$ as the concatenation of all binary variables that govern collision avoidance constraints for player $\nu$. In addition, we introduce $h = [h^\nu]_{\nu \in \mathcal{A}}$ and $\sigma = [\sigma^\nu]_{\nu \in \mathcal{A}} $ as the concatenation of all homotopy class variables and binary collision constraints flags, respectively. Finally, we denote $z^\nu := (x^\nu, u^\nu, h^\nu, \sigma^\nu)$ as the concatenation of all decision variables for player $\nu$, following the notation introduced in \cref{subsec:backgroundandproblemformulation}, and $z := [z^\nu]_{\nu \in \mathcal{A}} $ as the concatenation of the decision variables for all players. Then, the GNEP is formulated as the following OCP:

    \begin{equation}
    \label{eq:gnep2}
P^\nu \left\{
\begin{split}
&\underset{\substack{z^\nu \in \mathcal{Z}^\nu(z^{-\nu})}}{\min} \ J^\nu(z^\nu), & \\[2ex]
&\text{s.t } \ x_{0} = x_{\text{init}},
\end{split}
\right.
\end{equation}

where the feasible set $\mathcal{Z}^\nu$ is defined as follows:

\begin{equation}
\label{eq:feasibleset}
    \mathcal{Z}^\nu(z^{-\nu}) = \left\{ z^\nu \ \middle| \  \begin{array}{lll} 
\text{Dynamics:} & (\ref{eq:dynamics})  \\
\text{Collisions: } & (\ref{eq:collision},\ref{eq:collision2}) & \text{ s.t } \mu \in \mathcal{I}^\nu \\
\text{Monotonicity: } & (\ref{eq:monotonicity}) &  \text{ s.t } \mu \in \mathcal{I}^\nu \\
\text{Homotopic: } & (\ref{eq:logical}) &\text{ s.t } \mu \in \mathcal{I}^\nu \\
\text{Precedence: } & (\ref{eqn:precedence}) & \\
\text{State \& Control: } & (\ref{eq:stateandcontrolconstraints}) &  \\
x_{0} = x_{\text{init}} & & 
                                            \end{array} \right\}
\end{equation}

\subsubsection{Potential Function}

As demonstrated in \cref{def:wpg}, the following choice of potential function, $F(z)$, satisfies \cref{def:wpg} (for proof, refer to~\cite[Section 4]{britzelmeier2019numerical}) with $c^\nu = 1~\forall \nu \in \mathcal{A}$:
\begin{equation}
\label{eq:potentialfunction}
\begin{split}
F(z) &= \sum_{\nu \in \mathcal{A}}  J^\nu.\\
\end{split}
\end{equation}

Although players implicitly cooperate to prevent collisions in (\ref{eq:potentialfunction}), the game retains its non-cooperative nature, with vehicles remaining self-interested in minimizing their personal cost functions. Despite this, the dynamics of the game lean more towards cooperation than adversarial behavior~\cite{geiger2021learning}. The tendency towards cooperation is primarily driven from the inherent symmetry of collision costs; by avoiding a collision cost, a vehicle simultaneously helps in avoiding a collision cost for others.

\subsubsection{Overall Generalized Potential Game (GPG) Problem Formulation}

Thus, as established in \cref{subsec:backgroundandproblemformulation}, we can find a globally optimal Nash equilibrium strategy for all players, $z^\star$ by solving a single MIQP instead of solving the joint optimization problems $\{ P^\nu\}_{\nu \in \mathcal{A}}$, as follows:

    \begin{equation}
    \label{eq:hardformulation}
\left\{
\begin{split}
&\underset{\substack{z \in \mathcal{Z}}}{\min} \ F(z), & \\[2ex]
&\text{s.t } \ x_{0} = x_{\text{init}},
\end{split}
\right.
\end{equation}

where $\mathcal{Z}$ is analogous to (\ref{eq:feasibleset}), except that it collects the equality and inequality constraints for all players $\nu \in \mathcal{A}$. Note that since $F(z)$ is quadratic with all $P^\nu \ \forall \nu \in \mathcal{A}$ are positive semidefinite in (\ref{eq:costfunc}), and all constraints which govern set $\mathcal{Z}$ are linear, then solving (\ref{eq:hardformulation}) leads to a globally optimal strategy~\cite{7795555,esterle2020optimal}.



\section{Simulation Testing}
\label{sec:simulationresults}
We test our formulation in (\ref{eq:hardformulation}) in two distinct approaches. First, we enumerate all possible homotopy classes and then solve for the optimal Nash equilibrium trajectory within each one. Second, we solve for the optimal homotopy class and its optimal Nash equilibrium trajectory simultaneously. This approach enables us to assess whether our formulation effectively leads the solver to a globally optimal Nash equilibrium (without the need to iterate all homotopy classes) and to evaluate the associated computational costs with doing so.

Additionally, we compare the performance of our formulation with one that is free of the homotopic constraint described in (\ref{eq:logical}). More specifically, the homotopic constraint-free implementation is equivalent to that presented in (\ref{eq:hardformulation}); however, precedence (\ref{eqn:precedence}) constraints are omitted, and the homotopic constraints (\ref{eq:logical}) are replaced with the following one:

\begin{equation}
    \begin{split}
     \sigma^{\nu,\mu}_A (k) + \sigma^{\nu,\mu}_B (k) &+ \sigma^{\nu,\mu}_C (k)\\ 
     &+  \sigma^{\nu,\mu}_D (k) + \sigma^{\nu,\mu}_E(k) + \sigma^{\nu,\mu}_F(k)   =  1,
    \end{split}
    \label{eq:logical2}
\end{equation}

which enforces that any time-step $k$, one of the constraints is active, i.e the $(s^\nu(k), s^\mu(k))$ pair is outside the collisions area of \cref{fig:homotopy_diagram}, without introducing any homotopy variable $h^{\nu,\mu}$ to incorporate dependencies between the binary variables ($\sigma_\square^{\nu,\mu}$) that govern the collision avoidance. 

\subsection{Setup and Implementation Details}

\begin{figure}[h]
    \centering
    \includegraphics[width = \linewidth]{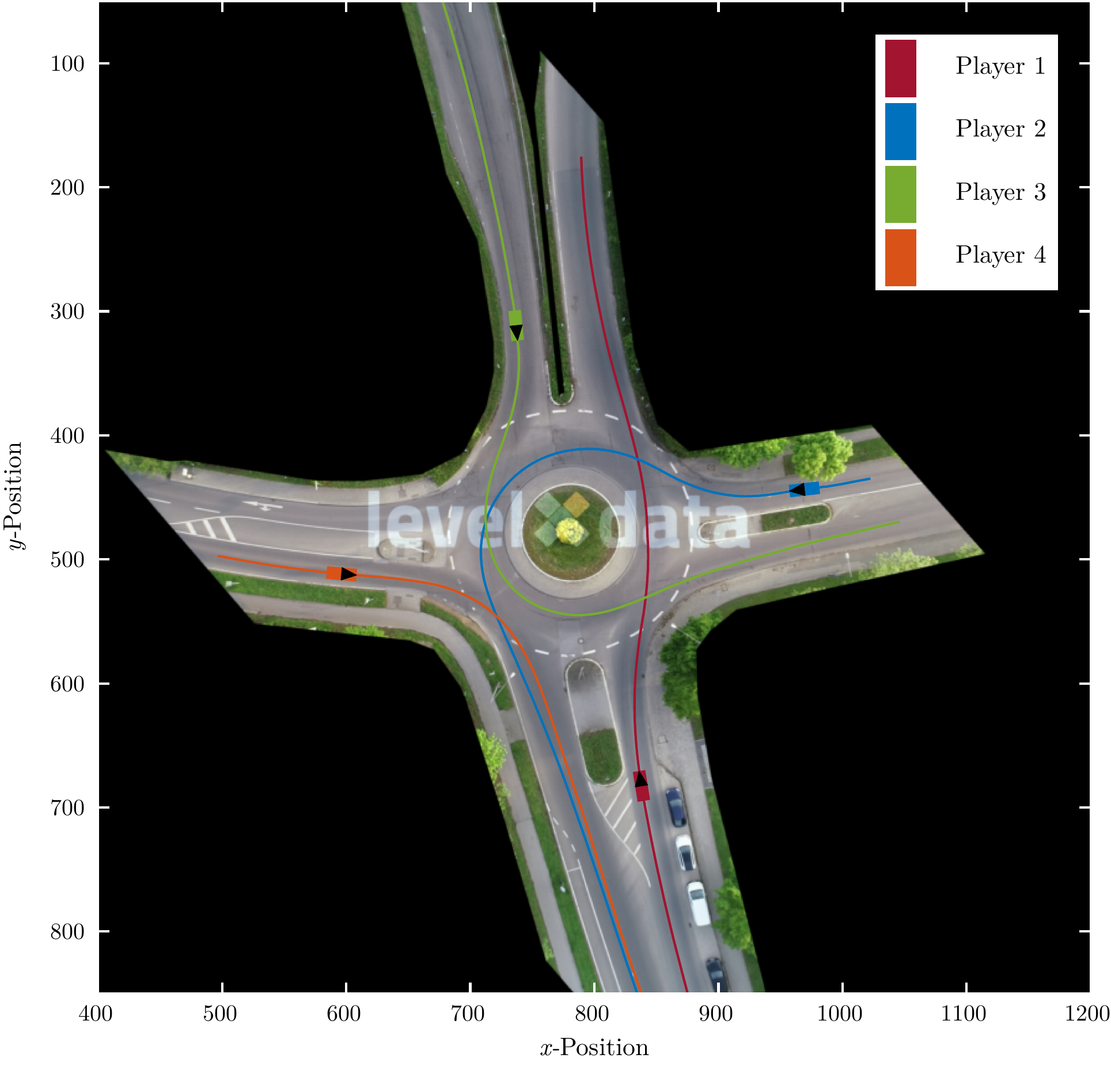}
    \caption{\textit{rounD} scenario used for this work~\cite{rounDdataset}. The scenario consists of 4 players, each with their predefined reference path, traversing Kackertstrasse roundabout.}
    \label{fig:results_scenario}
\end{figure} 
The problem is formulated in MATLAB~\cite{MATLAB} using YALMIP~\cite{Lofberg2004}\footnote{We incorporated custom low-level modifications to allow passing the initial guesses directly to the solver. See accompanying code for more details.} and solved using Gurobi\textsuperscript{\texttrademark}~\cite{gurobi} in a Model Predictive Control (MPC) fashion on a laptop with an Intel Core i7-11800H (2.30GHz) processor. We conduct the testing in a scenario extracted from \textit{rounD} dataset~\cite{rounDdataset}, as demonstrated in \cref{fig:results_scenario}. We consider a prediction horizon of $N = 35$ and time-step $dT = 0.1~\text{s}$.

We extract waypoints of naturalistic paths from four human-driven vehicles navigating around the Kackertstrasse roundabout, as shown in \cref{fig:results_scenario}. A parametrized reference path for each player is created by fitting a piecewise-continuous clothoid spline to the points using \texttt{referencePathFrenet} function~\cite{NavigationToolbox}. To obtain the parameters that characterize the conflict regions between the players, i.e $p^{\nu,\mu}_\dagger$ and $p^{\nu,\mu}_\ddagger$, as described in \cref{defconflicT}, we employed the following approach. Initially, the path, denoted as $r^\nu$, is discretized at a finite number of points where the global pose is explicitly defined, as demonstrated in \cref{fig:envelope3}. Subsequently, the vertices of the rectangle are computed. The envelope boundary and the relevant intersection points are computed using $\texttt{boundary}$ and $\texttt{polyxpoly}$ functions in MATLAB, respectively. Subsequently, $s^\nu_{\otimes1,\mu}$, $s^\nu_{\otimes2,\mu}$, $s^\mu_{\otimes1,\nu}$, and $s^\mu_{\otimes2,\nu}$ are obtained by using $\texttt{global2frenet}$ function. Finally, the parameters for the collisions area necessary for the problem formulation are then obtained using (\ref{eq:bounds}) and are reported in \cref{table:collisionareabounds} in the following format: $[\overline{s}^{\nu}_{\sqsubset,\mu}, \underline{s}^{\nu}_{\sqsubset,\mu}, \overline{s}^{\nu}_{\sqsupset,\mu}, \underline{s}^{\nu}_{\sqsupset,\mu}]$.
       
\begin{table}[htb!]
\resizebox{\columnwidth}{!}{%
\begin{tabular}{ccccc}
\toprule
 \diag{.1em}{1cm}{$\nu$}{$\mu$}   & Player 1                 & Player 2                 & Player 3                                        & Player 4                 \\ \midrule
Player 1 & \cellcolor[HTML]{333333} & $[74.8,79.1,79.3,86.6]$  & $[58.4,66.7,58.4,66.7]$                         & \cellcolor[HTML]{FE0000} \\ \hline
Player 2 & $[24.3,28.6,28.7,33.0]$  & \cellcolor[HTML]{333333} & $[43.3,47.7,56.4,60.8]$                         & $[56.6,60.9,-,-]$        \\ \hline
Player 3 & $[90.7,98.9,90.7,98.9]$  & $[60.8,65.2,73.7,78.0]$  & \cellcolor[HTML]{333333}{\color[HTML]{000000} } & \cellcolor[HTML]{FE0000} \\ \hline
Player 4 & \cellcolor[HTML]{FE0000} & $[28.6,32.9,-,-]$        & \cellcolor[HTML]{FE0000}                        & \cellcolor[HTML]{333333} \\ \bottomrule\\
\end{tabular}%
}
\caption{Collision area bounds in the following format $[\overline{s}^{\nu}_{\sqsubset,\mu}, \underline{s}^{\nu}_{\sqsubset,\mu}, \overline{s}^{\nu}_{\sqsupset,\mu}, \underline{s}^{\nu}_{\sqsupset,\mu}]$ which are used to construct the collision avoidance constraints (\ref{eq:collision}) and (\ref{eq:collision2})}
\label{table:collisionareabounds}
\end{table}

The length and width of all players is $3.6~\text{m}$ and $1.5~\text{m}$, respectively. The initial state of the players and their respective cost function parameters are reported in \cref{table:tableinitialstate}.

\begin{table}[htb!]
\centering
\begin{tabular}{ccc}
\toprule
Player $\nu$ & \begin{tabular}[c]{@{}c@{}}Initial state\\ $[s^\nu_0, \dot{s}^\nu_0]$\end{tabular} & \begin{tabular}[c]{@{}c@{}}Cost function parameters\\ $P^\nu, r^\nu$\end{tabular} \\
\midrule
1      & $[40, 2.5]$                                                                                                  & 1, 5                                                                              \\
2      & $[8, 3]$                                                                                                     & 1, 5                                                                              \\
3      & $[45,1]$                                                                                                     & 1, 5                                                                              \\
4      & $[15,3]$                                                                                                     & 1, 5                                                                                        \\
\bottomrule\\
\end{tabular}%
\caption{Players initial states and cost function parameters}
\label{table:tableinitialstate}
\end{table}

\subsection{Results and Discussion} 

\begin{table}[H]
\resizebox{\columnwidth}{!}{%
\midsepremove
\begin{tabular}{
  @{\extracolsep{\fill}}
c
c
c
c
c
c
  @{}
}
\toprule
\begin{tabular}[c]{@{}c@{}}Scenario Homotopy Class\\ $\langle h^{1,2}, h^{1,3}, h^{2,3}, h^{2,4} \rangle$\end{tabular} & \begin{tabular}[c]{@{}c@{}}NCT (s)\end{tabular} & \begin{tabular}[c]{@{}c@{}} TCT (s)\end{tabular} & \begin{tabular}[c]{@{}c@{}}NCE ($\text{m}/\text{s}^2$)\end{tabular} & \begin{tabular}[c]{@{}c@{}}NP (m)\end{tabular} & \begin{tabular}[c]{@{}c@{}}NP/TCT\\ ($\text{m}/\text{s}$)\end{tabular} \\
\specialrule{0.4pt}{0pt}{-10pt} \\
$\langle 0, 0, 0, 0 \rangle$                                        & 10.09        & 22.4      & 47.205        & 277.53        & 12.39           \\
$\langle 0, 0, 0, 1 \rangle$                                        & 7.59        & 22.4      & 52.068       & 318.76        & 14.23             \\
$\langle 0, 0, 1, 0 \rangle$                                        & 7.91        & 20.3      & 48.673       & 283.15        & 13.95        \\
$\langle 0, 0, 1, 1 \rangle$                                        & 4.89        & 15.6      & 49.913      & 294.79        & 18.90                 \\
$\langle 0, 1, 0, 0 \rangle$                                        & \multicolumn{5}{c}{\textbf{DEADLOCK}}                  \\
$\langle 0, 1, 0, 1 \rangle$                                        & \multicolumn{5}{c}{\textbf{DEADLOCK}}          \\
$\langle 0, 1, 1, 0 \rangle$                                        & 11.28        & 31.3      & 48.599       &  283.01       & 9.04                                \\
$\langle 0, 1, 1, 1 \rangle$                                        & 8.63        & 26.6      & 50.378       & 296.81        & 11.16                                                                \\
$\langle 1, 0, 0, 0 \rangle$                                        & 5.95        & 13.6      & 44.276      & 263.22        & 19.35                                                                \\
$\langle 1, 0, 0, 1 \rangle$                                        & 4.60        & 13.6      & 49.805        & 310.68        & 22.84                    \\
$\langle1, 0, 1, 0 \rangle$                                         & 6.23        & 13.0      & 46.442       & 271.25        & 20.87                                                  \\
\rowcolor[HTML]{DCDCDC}$\langle 1, 0, 1, 1 \rangle$                 & 3.25        & 10.4      & 39.309       & 265.13        & 25.49              \\
$\langle 1, 1, 0, 0 \rangle$                                        & 8.93        & 21.1      & 50.914       & 303.98       & 14.41                              \\
$\langle 1, 1, 0, 1 \rangle$                                        & 7.03        & 21.1      & 52.369      & 315.55        & 14.96                          \\
$\langle 1, 1, 1, 0 \rangle$                                        & 9.14        & 17.5      & 48.677       & 290.46        & 16.60                                                                \\
$\langle 1, 1, 1, 1 \rangle$                                        & 6.22        & 17.5      & 52.756       & 315.60        &  18.03                        \\ \hdashline[1pt/2pt]
No predefined homotopy                                              & 4.62        & 10.4      & 39.337        & 265.66        & 25.30                            \\ \hdashline[1pt/2pt]
Homotopic constraint-free                                                       & 15.83        & 10.4      & 39.301        & 265.13        & 25.49\\
\bottomrule \\
\end{tabular}%
}
\caption{Simulation results for the three test cases, separated by dashed lines: first, all 16 candidate scenario homotopy classes were iteratively tested; second, the solver was tasked with solving for the optimal homotopy and its trajectory without predefined constraints; third, the homotopic constraint (\ref{eq:logical}) is replaced with (\ref{eq:logical2}). The highlighted row shows the optimal scenario homotopy class. \textit{Legend for columns:} NCT: Net Computational Time, TCT: Task Completion Time, NCE: Net Control Effort, and NP: Net Progress.}
\label{table:prelim_results}
\end{table}

We iteratively tested the 16 candidate scenario homotopy classes, as outlined in \cref{table:prelim_results}. It is noteworthy that our solver successfully finds an optimal joint trajectory in all homotopy classes except for 2 cases, marked as \textit{deadlock}. For instance $\langle 0, 1, 0, 0\rangle$ represents a deadlock scenario as it requires Player 1 to precede Player 2, Player 3 to precede Player 1, and Player 2 to precede Player 3, an infeasible arrangement. We provide a methodology to identify deadlock cases in \cref{deadlock} a priori.

Additionally, we consider some evaluation metrics to assess the joint trajectory plan within each homotopy class: task completion time (TCT), net control effort (NCE), net progress (NP), and normalized net progress ($\frac{\text{NP}}{\text{TCT}}$). TCT quantifies the duration from the start of the simulation until all players exit the roundabout. NCE is computed by adding the 2-norm of the acceleration exerted by each vehicle throughout the task, which reflects the overall effort (or energy) in navigating the roundabout. NP denotes the collective progress made by all vehicles by the time all players exit the roundabout. The net progress is normalized by dividing it by the TCT to provide a more fair assessment of the quality of the homotopy class. 

It is evident that the performance varies significantly across the homotopy classes. For instance, while the TCT of the globally optimal homotopy class $\langle 1, 0, 1, 1 \rangle$ is around $10.4~\text{s}$, that of the worst-case scenario $\langle 0, 1, 1, 0 \rangle$ is roughly three times longer. Notably, the normalized progress of the optimal homotopy class is the highest, which is indicative of a fast progress rate to task completion.

Furthermore, we devised a solver capable of identifying the optimal scenario homotopy class and the optimal joint trajectory within it, denoted as \textit{No predefined homotopy} in \cref{table:prelim_results}. In this case, the optimal homotopy class is $\langle 1, 0, 1, 1 \rangle$, which aligns with the result obtained through the iterative testing of all 16 candidate scenario homotopy classes. It is essential to highlight that while both homotopy classes coincide, the optimal trajectories do not precisely align due to the finite length of the optimization horizon. Nonetheless, the discrepancy is very minimal, as evidenced by matching task completion times, with negligible differences observed in NP and NCE. It is important to highlight that we are able to obtain the globally optimal solution without explicitly enumerating all possible homotopy classes, achieving this in comparable computational time.

Finally, we present the results of a problem formulation where the homotopic constraint (\ref{eq:logical}) is replaced with (\ref{eq:logical2}), denoted as \textit{homotopic constraint-free}. It is important to note that while the passing sequence and other metrics are equivalent to the optimal one, the computational time required is approximately five times greater. While the homotopy-based implementation introduces additional binary variables ($h^{\nu,\mu}$), they create dependencies among previously independent binary variables($\sigma_\square^{\nu,\mu}$). This interdependence significantly aids the branch-and-bound solver in navigating the solution space more efficiently, as evidenced by the reduced computational time.

\begin{table}[H]
\resizebox{\columnwidth}{!}{%
\begin{tabular}{
  @{\extracolsep{\fill}}
ccccccccccccc
  @{}
}
\toprule
                                                                                                                                         & \multicolumn{3}{c}{\textit{Player 1}} & \multicolumn{3}{c}{\textit{Player 2}} & \multicolumn{3}{c}{\textit{Player 3}} & \multicolumn{3}{c}{\textit{Player 4}} \\
                                                                                                                                         \cmidrule(lr){2-4} \cmidrule(lr){5-7}\cmidrule(lr){8-10}\cmidrule(lr){11-13}
\multirow{-2}{*}{\begin{tabular}[c]{@{}c@{}}Scenario Homotopy Class\\ $\langle h^{1,2}, h^{1,3}, h^{2,3}, h^{2,4} \rangle$\end{tabular}} & $t^1_{r}$   & $c^1_{r}$  & $t^1_{w}$  & $t^2_{r}$   & $c^2_{r}$  & $t^2_{w}$  & $t^2_{r}$   &$c^2_{r}$  & $t^3_{w}$  & $t^2_{r}$   & $c^2_{r}$  & $t^4_{w}$  \\
\specialrule{0.4pt}{0pt}{-10pt} \\
$\langle 0, 0, 0, 0 \rangle$                                                                                                             & 3.9         & 5.5845     & 0.3        & 10.5        & 7.8700     & 1.3        & 18.5        & 8.8503     & 0.9     & 15.1        & 6.5674     & 10.8        \\
$\langle 0, 0, 0, 1 \rangle$                                                                                                             & 3.9         & 5.5845     & 0.3        & 10.5        & 7.8362     & 1.3        & 18.5        & 8.8518     & 0.9     & 2.3         & 4.1641     & 0.4        \\
$\langle 0, 0, 1, 0 \rangle$                                                                                                             & 3.9         & 5.5823     & 0.3        & 10.5        & 7.8649     & 1.3        & 6.8         & 6.8991      & 0.5     & 15.1        & 6.5674     & 10.8        \\
$\langle 0, 0, 1, 1 \rangle$                                                                                                             & 3.9         & 5.5958     & 0.3        & 10.5        & 7.7844     & 1.3        & 6.8         & 6.8994      & 0.5    & 2.3         & 4.1641     & 0.4        \\
$\langle 0, 1, 0, 0 \rangle$                                                                                                             & \multicolumn{12}{c}{\textbf{DEADLOCK}}                                                                                                                        \\
$\langle 0, 1, 0, 1 \rangle$                                                                                                             & \multicolumn{12}{c}{\textbf{DEADLOCK}}                                                                                                                        \\
$\langle 0, 1, 1, 0 \rangle$                                                                                                             & 11.8        & 7.9007     & 2.8        & 21.3        & 8.5864     & 1.3       & 6.8         & 7.0119        & 0.5     & 26.0        & 6.5674     & 21.7        \\
$\langle 0, 1, 1, 1 \rangle$                                                                                                             & 11.8        & 7.7874     & 2.8        & 21.4        & 8.5848     & 1.3        & 6.8         & 7.0123       &0.5     & 2.3         & 4.1641     & 0.4        \\
$\langle 1, 0, 0, 0 \rangle$                                                                                                             & 4.0         & 5.2306     & 0.3        & 4.8         & 6.7386     & 0.4        & 9.7         & 7.0248        & 0.8    & 7.7         & 6.2438     & 3.7        \\
$\langle 1, 0, 0, 1 \rangle$                                                                                                             & 4.0         & 5.2352     & 0.3        & 4.8         & 6.7055     & 0.4        & 9.7         & 7.0258         & 0.8     & 2.3         & 4.1641     & 0.4        \\
$\langle1, 0, 1, 0 \rangle$                                                                                                              & 4.0         & 5.2326     & 0.3        & 5.0         & 6.0411     & 0.4        & 6.8         & 7.0266        & 0.5     & 7.9         & 6.2628     & 3.9       \\
\rowcolor[HTML]{DCDCDC} 
$\langle 1, 0, 1, 1 \rangle$                                                                                                             & 4.0         & 5.2316     & 0.3        & 5.0         & 6.0040     & 0.4        & 6.8         &  7.0058         & 0.5     & 2.3         & 4.1641     & 0.4        \\
$\langle 1, 1, 0, 0 \rangle$                                                                                                             & 15.5        & 7.9159     & 2.7        & 4.8         & 6.7466     & 0.4        & 9.7         & 7.0297        &0.8    & 7.7         & 6.2434     & 3.7        \\
$\langle 1, 1, 0, 1 \rangle$                                                                                                             & 15.4        & 7.9195     & 2.7        & 4.8         & 6.7023     & 0.4        & 9.7         &  7.1546        & 0.8     & 2.3         & 4.1641     & 0.4        \\
$\langle 1, 1, 1, 0 \rangle$                                                                                                             & 11.8        & 7.7655     & 2.9        & 5.0         & 6.0515     & 0.4        & 6.8         & 7.0915        & 0.5    & 7.9         & 6.2628     &3.9        \\
$\langle 1, 1, 1, 1 \rangle$                                                                                                             & 11.8        & 7.7709     & 2.9        & 5.0         & 6.0108     & 0.4        & 6.8         & 7.0935        & 0.5   & 2.3         & 4.1641     & 0.4        \\ \hdashline[1pt/2pt]
No predefined homotopy                                                                                                                   & 4.0         & 5.2281     & 0.3        & 5.0         & 6.0172     & 0.4        & 6.8         & 7.0139        & 0.5     & 2.3         & 4.1641     & 0.4       \\ \hdashline[1pt/2pt]
Homotopy-free                                                                                                                    & 4.0         & 5.2328     & 0.3        & 5.0         & 6.0170     & 0.4        & 6.8         & 7.0078        & 0.5     & 2.3         & 4.2245     & 0.4       \\
\bottomrule \\
\end{tabular}%
}
\caption{Evaluation metrics for each player as they traverse the roundabout}
\label{table:results_2}
\end{table}

Additionally, in \cref{table:results_2}, we evaluate three additional scenario-specific metrics. $t^\nu_r$, $c^\nu_r$, represent the time and control effort player $\nu$ spends while navigating the roundabout, respectively. $t^\nu_w$ represents the time the vehicle spends within $1.5~\text{m}$ of the roundabout entry, indicative of the wait-time to enter the roundabout. It is evident that $\langle 1, 0, 1, 1 \rangle$ emerges as optimal from a social perspective~\cite{zanardi2021urban,la2016potential}, as it minimizes the cumulative time, control effort, and waiting time that vehicles spend navigating the roundabout. 

This opens the space to investigate some heuristics that could help determine \textit{good} homotopy candidates and \textit{bad} ones prior to sending them to the solver, thus saving computational effort. In fact, if our heuristics are good enough, we can find the global optimal trajectory by solving only one sub-problem, or at least set the value of some of the scenario homotopy class components. Additionally, priority-based heuristics could be employed to generate multi-agent motion plans where the precedence among (some) vehicles is determined a priori (for instance, it is established that an ambulance is priortized in traffic scenarios). More specifically, some components of a scenario homotopy class could be set while others are optimized using the framework.

\section{Conclusion}
\label{sec:conclusion}
In this work, we presented a homotopy-aware game-theoretic framework for joint trajectory prediction and optimization in urban traffic scenarios. Our approach allows us to effectively partition the high-dimensional solution space into simpler subregions, each corresponding to different high-level strategic decisions among the vehicles. The proposed formulation facilitates a comprehensive exploration of potential solutions and significantly enhances the computational efficiency of solving the trajectory planning problem (compared to a homotopy-free implementation). Through testing using scenarios extracted from \textit{rounD} dataset, the results have confirmed that the proposed model achieves a reduction in computational time and an increase in planning efficiency without compromising the quality of the solutions.

Future developments will focus on expanding the applicability of our framework to incorporate uncertainty on intent. We believe this could be addressed through the introduction of fictitious players. Additionally, a fruitful avenue could be giving the problem additional structure for cases where 3 or more players share the same conflict region, as it could lead to reducing the number of homotopy variables that are introduced as they are no longer independent.

\bibliographystyle{IEEEtran}
\bibliography{references.bib}

\begin{thebibliography}{100}
\providecommand{\url}[1]{#1}
\csname url@samestyle\endcsname
\providecommand{\newblock}{\relax}
\providecommand{\bibinfo}[2]{#2}
\providecommand{\BIBentrySTDinterwordspacing}{\spaceskip=0pt\relax}
\providecommand{\BIBentryALTinterwordstretchfactor}{4}
\providecommand{\BIBentryALTinterwordspacing}{\spaceskip=\fontdimen2\font plus
\BIBentryALTinterwordstretchfactor\fontdimen3\font minus \fontdimen4\font\relax}
\providecommand{\BIBforeignlanguage}[2]{{%
\expandafter\ifx\csname l@#1\endcsname\relax
\typeout{** WARNING: IEEEtran.bst: No hyphenation pattern has been}%
\typeout{** loaded for the language `#1'. Using the pattern for}%
\typeout{** the default language instead.}%
\else
\language=\csname l@#1\endcsname
\fi
#2}}
\providecommand{\BIBdecl}{\relax}
\BIBdecl

\bibitem{trautman2010unfreezing}
P.~Trautman and A.~Krause, ``Unfreezing the robot: Navigation in dense, interacting crowds,'' in \emph{2010 IEEE/RSJ International Conference on Intelligent Robots and Systems}.\hskip 1em plus 0.5em minus 0.4em\relax IEEE, 2010, pp. 797--803.

\bibitem{liebner2012driver}
M.~Liebner, M.~Baumann, F.~Klanner, and C.~Stiller, ``Driver intent inference at urban intersections using the intelligent driver model,'' in \emph{2012 IEEE intelligent vehicles symposium}.\hskip 1em plus 0.5em minus 0.4em\relax IEEE, 2012, pp. 1162--1167.

\bibitem{salzmann2020trajectron}
T.~Salzmann, B.~Ivanovic, P.~Chakravarty, and M.~Pavone, ``Trajectron++: Dynamically-feasible trajectory forecasting with heterogeneous data,'' in \emph{Computer Vision--ECCV 2020: 16th European Conference, Glasgow, UK, August 23--28, 2020, Proceedings, Part XVIII 16}.\hskip 1em plus 0.5em minus 0.4em\relax Springer, 2020, pp. 683--700.

\bibitem{bansal2018chauffeurnet}
M.~Bansal, A.~Krizhevsky, and A.~Ogale, ``Chauffeurnet: Learning to drive by imitating the best and synthesizing the worst,'' \emph{arXiv preprint arXiv:1812.03079}, 2018.

\bibitem{yuan2021agentformer}
Y.~Yuan, X.~Weng, Y.~Ou, and K.~M. Kitani, ``Agentformer: Agent-aware transformers for socio-temporal multi-agent forecasting,'' in \emph{Proceedings of the IEEE/CVF International Conference on Computer Vision}, 2021, pp. 9813--9823.

\bibitem{pmlr-v100-chai20a}
\BIBentryALTinterwordspacing
Y.~Chai, B.~Sapp, M.~Bansal, and D.~Anguelov, ``Multipath: Multiple probabilistic anchor trajectory hypotheses for behavior prediction,'' in \emph{Proceedings of the Conference on Robot Learning}, ser. Proceedings of Machine Learning Research, L.~P. Kaelbling, D.~Kragic, and K.~Sugiura, Eds., vol. 100.\hskip 1em plus 0.5em minus 0.4em\relax PMLR, 30 Oct--01 Nov 2020, pp. 86--99. [Online]. Available: \url{https://proceedings.mlr.press/v100/chai20a.html}
\BIBentrySTDinterwordspacing

\bibitem{9526613}
W.~Ding, L.~Zhang, J.~Chen, and S.~Shen, ``Epsilon: An efficient planning system for automated vehicles in highly interactive environments,'' \emph{IEEE Transactions on Robotics}, vol.~38, no.~2, pp. 1118--1138, 2022.

\bibitem{8569729}
C.~Hubmann, J.~Schulz, G.~Xu, D.~Althoff, and C.~Stiller, ``A belief state planner for interactive merge maneuvers in congested traffic,'' in \emph{2018 21st International Conference on Intelligent Transportation Systems (ITSC)}, 2018, pp. 1617--1624.

\bibitem{7795596}
N.~Evestedt, E.~Ward, J.~Folkesson, and D.~Axehill, ``Interaction aware trajectory planning for merge scenarios in congested traffic situations,'' in \emph{2016 IEEE 19th International Conference on Intelligent Transportation Systems (ITSC)}, 2016, pp. 465--472.

\bibitem{le2022algames}
S.~Le~Cleac’h, M.~Schwager, and Z.~Manchester, ``{ALGAMES: a fast augmented Lagrangian solver for constrained dynamic games},'' \emph{Autonomous Robots}, vol.~46, no.~1, pp. 201--215, 2022.

\bibitem{zanardi2021urban}
A.~Zanardi, E.~Mion, M.~Bruschetta, S.~Bolognani, A.~Censi, and E.~Frazzoli, ``{Urban Driving Games With Lexicographic Preferences and Socially Efficient Nash Equilibria},'' \emph{IEEE Robotics and Automation Letters}, vol.~6, no.~3, pp. 4978--4985, 2021.

\bibitem{zhu2022sequential}
E.~L. Zhu and F.~Borrelli, ``{A Sequential Quadratic Programming Approach to the Solution of Open-Loop Generalized Nash Equilibria},'' in \emph{2023 IEEE International Conference on Robotics and Automation (ICRA)}, 2023, pp. 3211--3217.

\bibitem{evens2022learning}
B.~Evens, M.~Schuurmans, and P.~Patrinos, ``{Learning MPC for Interaction-Aware Autonomous Driving: A Game-Theoretic Approach},'' in \emph{2022 European Control Conference (ECC)}, 2022, pp. 34--39.

\bibitem{fridovich2020efficient}
D.~Fridovich-Keil, E.~Ratner, L.~Peters, A.~D. Dragan, and C.~J. Tomlin, ``{Efficient Iterative Linear-Quadratic Approximations for Nonlinear Multi-Player General-Sum Differential Games},'' in \emph{2020 IEEE International Conference on Robotics and Automation (ICRA)}, 2020, pp. 1475--1481.

\bibitem{britzelmeier2019numerical}
A.~Britzelmeier, A.~Dreves, and M.~Gerdts, ``{Numerical solution of potential games arising in the control of cooperative automatic vehicles},'' in \emph{2019 Proceedings of the conference on control and its applications}.\hskip 1em plus 0.5em minus 0.4em\relax SIAM, 2019, pp. 38--45.

\bibitem{laine2023computation}
F.~Laine, D.~Fridovich-Keil, C.-Y. Chiu, and C.~Tomlin, ``{The computation of approximate generalized feedback nash equilibria},'' \emph{SIAM Journal on Optimization}, vol.~33, no.~1, pp. 294--318, 2023.

\bibitem{dreves2019best}
A.~Dreves, ``A best-response approach for equilibrium selection in two-player generalized nash equilibrium problems,'' \emph{Optimization}, vol.~68, no.~12, pp. 2269--2295, 2019.

\bibitem{bhattacharya_2021}
\BIBentryALTinterwordspacing
S.~Bhattacharya, M.~Likhachev, and V.~Kumar, ``Search-based path planning with homotopy class constraints in 3d,'' \emph{Proceedings of the AAAI Conference on Artificial Intelligence}, vol.~26, no.~1, pp. 2097--2099, Sep. 2021. [Online]. Available: \url{https://ojs.aaai.org/index.php/AAAI/article/view/8435}
\BIBentrySTDinterwordspacing

\bibitem{chen2023interactive}
Y.~Chen, S.~Veer, P.~Karkus, and M.~Pavone, ``Interactive joint planning for autonomous vehicles,'' \emph{IEEE Robotics and Automation Letters}, 2023.

\bibitem{facchinei2010generalized}
F.~Facchinei and C.~Kanzow, ``{Generalized Nash equilibrium Problems},'' \emph{Annals of Operations Research}, vol. 175, no.~1, pp. 177--211, 2010.

\bibitem{10460127}
M.~Khayyat, S.~Bolognani, S.~Arrigoni, and F.~Braghin, ``A sampling-based approach to urban motion planning games with stochastic dynamics,'' \emph{IEEE Transactions on Intelligent Vehicles}, pp. 1--15, 2024.

\bibitem{facchinei2011decomposition}
F.~Facchinei, V.~Piccialli, and M.~Sciandrone, ``{Decomposition Algorithms for Generalized Potential Games},'' \emph{Computational Optimization and Applications}, vol.~50, pp. 237--262, 2011.

\bibitem{sagratella2017algorithms}
S.~Sagratella, ``Algorithms for generalized potential games with mixed-integer variables,'' \emph{Computational Optimization and Applications}, vol.~68, no.~3, pp. 689--717, 2017.

\bibitem{rounDdataset}
R.~Krajewski, T.~Moers, J.~Bock, L.~Vater, and L.~Eckstein, ``The round dataset: A drone dataset of road user trajectories at roundabouts in germany,'' in \emph{2020 IEEE 23rd International Conference on Intelligent Transportation Systems (ITSC)}, 2020, pp. 1--6.

\bibitem{bacsar1998dynamic}
T.~Ba{\c{s}}ar and G.~J. Olsder, \emph{Dynamic noncooperative game theory}.\hskip 1em plus 0.5em minus 0.4em\relax SIAM, 1998.

\bibitem{leyton2022essentials}
K.~Leyton-Brown and Y.~Shoham, \emph{{Essentials of game theory: A concise multidisciplinary introduction}}.\hskip 1em plus 0.5em minus 0.4em\relax Springer Nature, 2022.

\bibitem{fischer2014generalized}
A.~Fischer, M.~Herrich, and K.~Sch{\"o}nefeld, ``{Generalized Nash equilibrium Problems-Recent Advances and Challenges},'' \emph{Pesquisa Operacional}, vol.~34, pp. 521--558, 2014.

\bibitem{9341137}
M.~Wang, N.~Mehr, A.~Gaidon, and M.~Schwager, ``Game-theoretic planning for risk-aware interactive agents,'' in \emph{2020 IEEE/RSJ International Conference on Intelligent Robots and Systems (IROS)}, 2020, pp. 6998--7005.

\bibitem{fisac2019hierarchical}
J.~F. Fisac, E.~Bronstein, E.~Stefansson, D.~Sadigh, S.~S. Sastry, and A.~D. Dragan, ``Hierarchical game-theoretic planning for autonomous vehicles,'' in \emph{2019 International conference on robotics and automation (ICRA)}.\hskip 1em plus 0.5em minus 0.4em\relax IEEE, 2019, pp. 9590--9596.

\bibitem{williams2016aggressive}
G.~Williams, P.~Drews, B.~Goldfain, J.~M. Rehg, and E.~A. Theodorou, ``{Aggressive driving with model predictive path integral control},'' in \emph{2016 IEEE International Conference on Robotics and Automation (ICRA)}, 2016, pp. 1433--1440.

\bibitem{spica2020real}
R.~Spica, E.~Cristofalo, Z.~Wang, E.~Montijano, and M.~Schwager, ``{A real-time game theoretic planner for autonomous two-player drone racing},'' \emph{IEEE Transactions on Robotics}, vol.~36, no.~5, pp. 1389--1403, 2020.

\bibitem{wang2021game}
M.~Wang, Z.~Wang, J.~Talbot, J.~C. Gerdes, and M.~Schwager, ``Game-theoretic planning for self-driving cars in multivehicle competitive scenarios,'' \emph{IEEE Transactions on Robotics}, vol.~37, no.~4, pp. 1313--1325, 2021.

\bibitem{liniger2019noncooperative}
A.~Liniger and J.~Lygeros, ``{A noncooperative game approach to autonomous racing},'' \emph{IEEE Transactions on Control Systems Technology}, vol.~28, no.~3, pp. 884--897, 2019.

\bibitem{notomista2020enhancing}
G.~Notomista, M.~Wang, M.~Schwager, and M.~Egerstedt, ``Enhancing game-theoretic autonomous car racing using control barrier functions,'' in \emph{2020 IEEE international conference on robotics and automation (ICRA)}.\hskip 1em plus 0.5em minus 0.4em\relax IEEE, 2020, pp. 5393--5399.

\bibitem{9992957}
K.~Miller and S.~Mitra, ``Multi-agent motion planning using differential games with lexicographic preferences,'' in \emph{2022 IEEE 61st Conference on Decision and Control (CDC)}, 2022, pp. 5751--5756.

\bibitem{10178143}
R.~Reiter, J.~Hoffmann, J.~Boedecker, and M.~Diehl, ``A hierarchical approach for strategic motion planning in autonomous racing,'' in \emph{2023 European Control Conference (ECC)}, 2023, pp. 1--8.

\bibitem{fabiani2019multi}
F.~Fabiani and S.~Grammatico, ``{Multi-vehicle automated driving as a generalized mixed-integer potential game},'' \emph{IEEE Transactions on Intelligent Transportation Systems}, vol.~21, no.~3, pp. 1064--1073, 2019.

\bibitem{singh2023globally}
A.~Singh and D.~Ghosh, ``A globally convergent improved bfgs method for generalized nash equilibrium problems,'' \emph{SeMA Journal}, pp. 1--27, 2023.

\bibitem{monderer1996potential}
D.~Monderer and L.~S. Shapley, ``{Potential games},'' \emph{Games and economic behavior}, vol.~14, no.~1, pp. 124--143, 1996.

\bibitem{la2016potential}
Q.~D. La, Y.~H. Chew, B.-H. Soong, Q.~D. L{\~a}, Y.~H. Chew, and B.-H. Soong, ``Potential games,'' \emph{Potential Game Theory: Applications in Radio Resource Allocation}, pp. 23--69, 2016.

\bibitem{carbonell2014refinements}
O.~Carbonell-Nicolau and R.~P. McLean, ``Refinements of nash equilibrium in potential games,'' \emph{Theoretical Economics}, vol.~9, no.~3, pp. 555--582, 2014.

\bibitem{multiplenash}
L.~Peters, D.~Fridovich-Keil, C.~J. Tomlin, and Z.~N. Sunberg, ``{Inference-Based Strategy Alignment for General-Sum Differential Games},'' in \emph{Proceedings of the 19th International Conference on Autonomous Agents and MultiAgent Systems}, ser. AAMAS '20.\hskip 1em plus 0.5em minus 0.4em\relax Richland, SC: International Foundation for Autonomous Agents and Multiagent Systems, 2020, p. 1037–1045.

\bibitem{britzelmeier2021decomposition}
A.~Britzelmeier and A.~Dreves, ``A decomposition algorithm for nash equilibria in intersection management,'' \emph{Optimization}, vol.~70, no.~11, pp. 2441--2478, 2021.

\bibitem{katriniok2017distributed}
A.~Katriniok, P.~Kleibaum, and M.~Jo{\v{s}}evski, ``Distributed model predictive control for intersection automation using a parallelized optimization approach,'' \emph{IFAC-PapersOnLine}, vol.~50, no.~1, pp. 5940--5946, 2017.

\bibitem{8263811}
D.~Dueri, Y.~Mao, Z.~Mian, J.~Ding, and B.~A{\c{c}}{\i}kme{\c{s}}e, ``Trajectory optimization with inter-sample obstacle avoidance via successive convexification,'' in \emph{2017 IEEE 56th Annual Conference on Decision and Control (CDC)}, 2017, pp. 1150--1156.

\bibitem{liu2018convex}
C.~Liu, C.-Y. Lin, and M.~Tomizuka, ``The convex feasible set algorithm for real time optimization in motion planning,'' \emph{SIAM Journal on Control and optimization}, vol.~56, no.~4, pp. 2712--2733, 2018.

\bibitem{zhang2022receding}
X.~Zhang, Y.~Jiang, Y.~Lu, and X.~Xu, ``Receding-horizon reinforcement learning approach for kinodynamic motion planning of autonomous vehicles,'' \emph{IEEE Transactions on Intelligent Vehicles}, vol.~7, no.~3, pp. 556--568, 2022.

\bibitem{schulman2014motion}
J.~Schulman, Y.~Duan, J.~Ho, A.~Lee, I.~Awwal, H.~Bradlow, J.~Pan, S.~Patil, K.~Goldberg, and P.~Abbeel, ``Motion planning with sequential convex optimization and convex collision checking,'' \emph{The International Journal of Robotics Research}, vol.~33, no.~9, pp. 1251--1270, 2014.

\bibitem{schulman2013finding}
J.~Schulman, J.~Ho, A.~X. Lee, I.~Awwal, H.~Bradlow, and P.~Abbeel, ``{Finding Locally Optimal, Collision-Free Trajectories with Sequential Convex Optimization},'' in \emph{Robotics: Science and Systems}, vol.~9, no.~1.\hskip 1em plus 0.5em minus 0.4em\relax Berlin, Germany, 2013, pp. 1--10.

\bibitem{mazumdar2019finding}
E.~V. Mazumdar, M.~I. Jordan, and S.~S. Sastry, ``On finding local nash equilibria (and only local nash equilibria) in zero-sum games,'' \emph{arXiv preprint arXiv:1901.00838}, 2019.

\bibitem{quirynen2023real}
R.~Quirynen, S.~Safaoui, and S.~Di~Cairano, ``Real-time mixed-integer quadratic programming for vehicle decision making and motion planning,'' \emph{arXiv preprint arXiv:2308.10069}, 2023.

\bibitem{schouwenaars2002safe}
T.~Schouwenaars, {\'E}.~F{\'e}ron, and J.~How, ``Safe receding horizon path planning for autonomous vehicles,'' in \emph{Proceedings of the Annual Allerton Conference on Communication Control and Computing}, vol.~40, no.~1.\hskip 1em plus 0.5em minus 0.4em\relax The University; 1998, 2002, pp. 295--304.

\bibitem{7795555}
X.~Qian, F.~Altché, P.~Bender, C.~Stiller, and A.~de~La~Fortelle, ``Optimal trajectory planning for autonomous driving integrating logical constraints: An miqp perspective,'' in \emph{2016 IEEE 19th International Conference on Intelligent Transportation Systems (ITSC)}, 2016, pp. 205--210.

\bibitem{schouwenaars2001mixed}
T.~Schouwenaars, B.~De~Moor, E.~Feron, and J.~How, ``Mixed integer programming for multi-vehicle path planning,'' in \emph{2001 European control conference (ECC)}.\hskip 1em plus 0.5em minus 0.4em\relax IEEE, 2001, pp. 2603--2608.

\bibitem{gurobi}
\BIBentryALTinterwordspacing
{Gurobi Optimization, LLC}, ``{Gurobi Optimizer Reference Manual},'' 2023. [Online]. Available: \url{https://www.gurobi.com}
\BIBentrySTDinterwordspacing

\bibitem{quirynen2023tailored}
R.~Quirynen and S.~Di~Cairano, ``Tailored presolve techniques in branch-and-bound method for fast mixed-integer optimal control applications,'' \emph{Optimal Control Applications and Methods}, vol.~44, no.~6, pp. 3139--3167, 2023.

\bibitem{7535369}
F.~Altché and A.~de~La~Fortelle, ``Analysis of optimal solutions to robot coordination problems to improve autonomous intersection management policies,'' in \emph{2016 IEEE Intelligent Vehicles Symposium (IV)}, 2016, pp. 86--91.

\bibitem{10.1145/3566097.3567849}
\BIBentryALTinterwordspacing
P.-C. Chen, X.~Liu, C.-W. Lin, C.~Huang, and Q.~Zhu, ``Mixed-traffic intersection management utilizing connected and autonomous vehicles as traffic regulators,'' in \emph{Proceedings of the 28th Asia and South Pacific Design Automation Conference}, ser. ASPDAC '23.\hskip 1em plus 0.5em minus 0.4em\relax New York, NY, USA: Association for Computing Machinery, 2023, p. 52–57. [Online]. Available: \url{https://doi.org/10.1145/3566097.3567849}
\BIBentrySTDinterwordspacing

\bibitem{9667332}
W.~Wu, Y.~Liu, W.~Liu, F.~Zhang, V.~Dixit, and S.~T. Waller, ``Autonomous intersection management for connected and automated vehicles: A lane-based method,'' \emph{IEEE Transactions on Intelligent Transportation Systems}, vol.~23, no.~9, pp. 15\,091--15\,106, 2022.

\bibitem{10178275}
A.~Caregnato-Neto, M.~R. O.~A. Maximo, and R.~J.~M. Afonso, ``A novel line of sight constraint for mixed-integer programming models with applications to multi-agent motion planning,'' in \emph{2023 European Control Conference (ECC)}, 2023, pp. 1--6.

\bibitem{10156567}
V.~Bhattacharyya and A.~Vahidi, ``Automated vehicle highway merging: Motion planning via adaptive interactive mixed-integer mpc,'' in \emph{2023 American Control Conference (ACC)}, 2023, pp. 1141--1146.

\bibitem{kessler2020linear}
T.~Kessler, K.~Esterle, and A.~Knoll, ``Linear differential games for cooperative behavior planning of autonomous vehicles using mixed-integer programming,'' in \emph{2020 59th IEEE Conference on Decision and Control (CDC)}.\hskip 1em plus 0.5em minus 0.4em\relax IEEE, 2020, pp. 4060--4066.

\bibitem{8814060}
T.~Kessler and A.~Knoll, ``Cooperative multi-vehicle behavior coordination for autonomous driving,'' in \emph{2019 IEEE Intelligent Vehicles Symposium (IV)}, 2019, pp. 1953--1960.

\bibitem{7995795}
J.~Eilbrecht and O.~Stursberg, ``Cooperative driving using a hierarchy of mixed-integer programming and tracking control,'' in \emph{2017 IEEE Intelligent Vehicles Symposium (IV)}, 2017, pp. 673--678.

\bibitem{ioan2021mixed}
D.~Ioan, I.~Prodan, S.~Olaru, F.~Stoican, and S.-I. Niculescu, ``Mixed-integer programming in motion planning,'' \emph{Annual Reviews in Control}, vol.~51, pp. 65--87, 2021.

\bibitem{hopcroft1984complexity}
J.~E. Hopcroft, J.~T. Schwartz, and M.~Sharir, ``On the complexity of motion planning for multiple independent objects; pspace-hardness of the" warehouseman's problem",'' \emph{The international journal of robotics research}, vol.~3, no.~4, pp. 76--88, 1984.

\bibitem{10384147}
A.~Zanardi, P.~Zullo, A.~Censi, and E.~Frazzoli, ``Factorization of multi-agent sampling-based motion planning,'' in \emph{2023 62nd IEEE Conference on Decision and Control (CDC)}, 2023, pp. 8108--8115.

\bibitem{9981416}
A.~Zanardi, S.~Bolognani, A.~Censi, F.~Dorfler, and E.~Frazzoli, ``{Factorization of Dynamic Games over Spatio-Temporal Resources},'' in \emph{2022 IEEE/RSJ International Conference on Intelligent Robots and Systems (IROS)}, 2022, pp. 13\,159--13\,166.

\bibitem{bhattacharya_2011}
S.~Bhattacharya, M.~Likhachev, and V.~Kumar, ``Identification and representation of homotopy classes of trajectories for search-based path planning in 3d,'' in \emph{Proceedings of Robotics: Science and Systems}, Los Angeles, CA, USA, June 2011.

\bibitem{7182335}
J.~Park, S.~Karumanchi, and K.~Iagnemma, ``Homotopy-based divide-and-conquer strategy for optimal trajectory planning via mixed-integer programming,'' \emph{IEEE Transactions on Robotics}, vol.~31, no.~5, pp. 1101--1115, 2015.

\bibitem{kant1986toward}
K.~Kant and S.~W. Zucker, ``Toward efficient trajectory planning: The path-velocity decomposition,'' \emph{The international journal of robotics research}, vol.~5, no.~3, pp. 72--89, 1986.

\bibitem{gregoire2014priority}
J.~Gregoire, ``Priority-based coordination of mobile robots,'' \emph{arXiv preprint arXiv:1410.0879}, 2014.

\bibitem{8574950}
B.~Yi, P.~Bender, F.~Bonarens, and C.~Stiller, ``Model predictive trajectory planning for automated driving,'' \emph{IEEE Transactions on Intelligent Vehicles}, vol.~4, no.~1, pp. 24--38, 2019.

\bibitem{8917274}
M.~Schmidt, C.~Wissing, J.~Braun, T.~Nattermann, and T.~Bertram, ``Maneuver identification for interaction-aware highway lane change behavior planning based on polygon clipping and convex optimization,'' in \emph{2019 IEEE Intelligent Transportation Systems Conference (ITSC)}, 2019, pp. 3948--3953.

\bibitem{8814241}
K.~Esterle, V.~Aravantinos, and A.~Knoll, ``From specifications to behavior: Maneuver verification in a semantic state space,'' in \emph{2019 IEEE Intelligent Vehicles Symposium (IV)}, 2019, pp. 2140--2147.

\bibitem{schafer2021computation}
L.~Sch{\"a}fer, S.~Manzinger, and M.~Althoff, ``Computation of solution spaces for optimization-based trajectory planning,'' \emph{IEEE Transactions on Intelligent Vehicles}, 2021.

\bibitem{7225909}
P.~Bender, O.~S. Tas, J.~Ziegler, and C.~Stiller, ``The combinatorial aspect of motion planning: Maneuver variants in structured environments,'' in \emph{2015 IEEE Intelligent Vehicles Symposium (IV)}, 2015, pp. 1386--1392.

\bibitem{gupta2023game}
K.~Gupta and D.~Fridovich-Keil, ``Game-theoretic occlusion-aware motion planning: an efficient hybrid-information approach,'' \emph{arXiv preprint arXiv:2309.10901}, 2023.

\bibitem{shapley1953stochastic}
L.~S. Shapley, ``{Stochastic games},'' \emph{Proceedings of the national academy of sciences}, vol.~39, no.~10, pp. 1095--1100, 1953.

\bibitem{zhan2017spatially}
W.~Zhan, J.~Chen, C.-Y. Chan, C.~Liu, and M.~Tomizuka, ``Spatially-partitioned environmental representation and planning architecture for on-road autonomous driving,'' in \emph{2017 IEEE Intelligent Vehicles Symposium (IV)}.\hskip 1em plus 0.5em minus 0.4em\relax IEEE, 2017, pp. 632--639.

\bibitem{mendelson1990introduction}
B.~Mendelson, \emph{Introduction to topology}.\hskip 1em plus 0.5em minus 0.4em\relax Courier Corporation, 1990.

\bibitem{nocedal1999numerical}
J.~Nocedal and S.~J. Wright, \emph{Numerical optimization}.\hskip 1em plus 0.5em minus 0.4em\relax Springer, 1999.

\bibitem{richards2015inter}
A.~Richards and O.~Turnbull, ``Inter-sample avoidance in trajectory optimizers using mixed-integer linear programming,'' \emph{International Journal of Robust and Nonlinear Control}, vol.~25, no.~4, pp. 521--526, 2015.

\bibitem{bertolazzi2015g1}
E.~Bertolazzi and M.~Frego, ``G1 fitting with clothoids,'' \emph{Mathematical Methods in the Applied Sciences}, vol.~38, no.~5, pp. 881--897, 2015.

\bibitem{silva2018clothoid}
J.~A. Silva and V.~Grassi, ``Clothoid-based global path planning for autonomous vehicles in urban scenarios,'' in \emph{2018 IEEE international conference on robotics and automation (ICRA)}.\hskip 1em plus 0.5em minus 0.4em\relax IEEE, 2018, pp. 4312--4318.

\bibitem{werling2010optimal}
M.~Werling, J.~Ziegler, S.~Kammel, and S.~Thrun, ``Optimal trajectory generation for dynamic street scenarios in a frenet frame,'' in \emph{2010 IEEE international conference on robotics and automation}.\hskip 1em plus 0.5em minus 0.4em\relax IEEE, 2010, pp. 987--993.

\bibitem{mercy2017spline}
T.~Mercy, R.~Van~Parys, and G.~Pipeleers, ``Spline-based motion planning for autonomous guided vehicles in a dynamic environment,'' \emph{IEEE Transactions on Control Systems Technology}, vol.~26, no.~6, pp. 2182--2189, 2017.

\bibitem{9951715}
S.~Arrigoni, M.~Pirovano, S.~Mentasti, M.~Khayyat, F.~Braghin, F.~Cheli, and M.~Matteucci, ``Experimental implementation of a trajectory planner for autonomous driving,'' in \emph{2022 AEIT International Annual Conference (AEIT)}, 2022, pp. 1--6.

\bibitem{kloeser2020nmpc}
D.~Kloeser, T.~Schoels, T.~Sartor, A.~Zanelli, G.~Prison, and M.~Diehl, ``Nmpc for racing using a singularity-free path-parametric model with obstacle avoidance,'' \emph{IFAC-PapersOnLine}, vol.~53, no.~2, pp. 14\,324--14\,329, 2020.

\bibitem{reiter2021parameterization}
R.~Reiter and M.~Diehl, ``Parameterization approach of the frenet transformation for model predictive control of autonomous vehicles,'' in \emph{2021 European Control Conference (ECC)}.\hskip 1em plus 0.5em minus 0.4em\relax IEEE, 2021, pp. 2414--2419.

\bibitem{9341731}
J.~L. Vázquez, M.~Brühlmeier, A.~Liniger, A.~Rupenyan, and J.~Lygeros, ``Optimization-based hierarchical motion planning for autonomous racing,'' in \emph{2020 IEEE/RSJ International Conference on Intelligent Robots and Systems (IROS)}, 2020, pp. 2397--2403.

\bibitem{novi2019real}
T.~Novi, A.~Liniger, R.~Capitani, and C.~Annicchiarico, ``Real-time control for at-limit handling driving on a predefined path,'' \emph{Vehicle system dynamics}, 2019.

\bibitem{chen2020mpc}
S.~Chen and H.~Chen, ``Mpc-based path tracking with pid speed control for autonomous vehicles,'' in \emph{IOP Conference Series: Materials Science and Engineering}, vol. 892, no.~1.\hskip 1em plus 0.5em minus 0.4em\relax IOP Publishing, 2020, p. 012034.

\bibitem{7810277}
R.~Verschueren, M.~Zanon, R.~Quirynen, and M.~Diehl, ``Time-optimal race car driving using an online exact hessian based nonlinear mpc algorithm,'' in \emph{2016 European Control Conference (ECC)}, 2016, pp. 141--147.

\bibitem{esterle2020optimal}
K.~Esterle, T.~Kessler, and A.~Knoll, ``Optimal behavior planning for autonomous driving: A generic mixed-integer formulation,'' in \emph{2020 IEEE Intelligent Vehicles Symposium (IV)}.\hskip 1em plus 0.5em minus 0.4em\relax IEEE, 2020, pp. 1914--1921.

\bibitem{geiger2021learning}
P.~Geiger and C.-N. Straehle, ``{Learning game-theoretic models of multiagent trajectories using implicit layers},'' in \emph{Proceedings of the AAAI Conference on Artificial Intelligence}, vol.~35, no.~6, 2021, pp. 4950--4958.

\bibitem{MATLAB}
\BIBentryALTinterwordspacing
T.~M. Inc., ``Matlab version: 9.13.0 (r2022b),'' Natick, Massachusetts, United States, 2022. [Online]. Available: \url{https://www.mathworks.com}
\BIBentrySTDinterwordspacing

\bibitem{Lofberg2004}
J.~L{\"{o}}fberg, ``Yalmip : A toolbox for modeling and optimization in matlab,'' in \emph{In Proceedings of the CACSD Conference}, Taipei, Taiwan, 2004.

\bibitem{NavigationToolbox}
\BIBentryALTinterwordspacing
T.~M. Inc., ``Navigation toolbox version: 2.20 (r2022b),'' Natick, Massachusetts, United States, 2022. [Online]. Available: \url{https://www.mathworks.com}
\BIBentrySTDinterwordspacing

\bibitem{bemporad2001discrete}
A.~Bemporad, F.~D. Torrisi, and M.~Morari, ``Discrete-time hybrid modeling and verification of the batch evaporator process benchmark,'' \emph{European Journal of Control}, vol.~7, no.~4, pp. 382--399, 2001.

\end{thebibliography}

\begin{IEEEbiography}[{\includegraphics[width=1in,height=1.25in,clip,keepaspectratio]{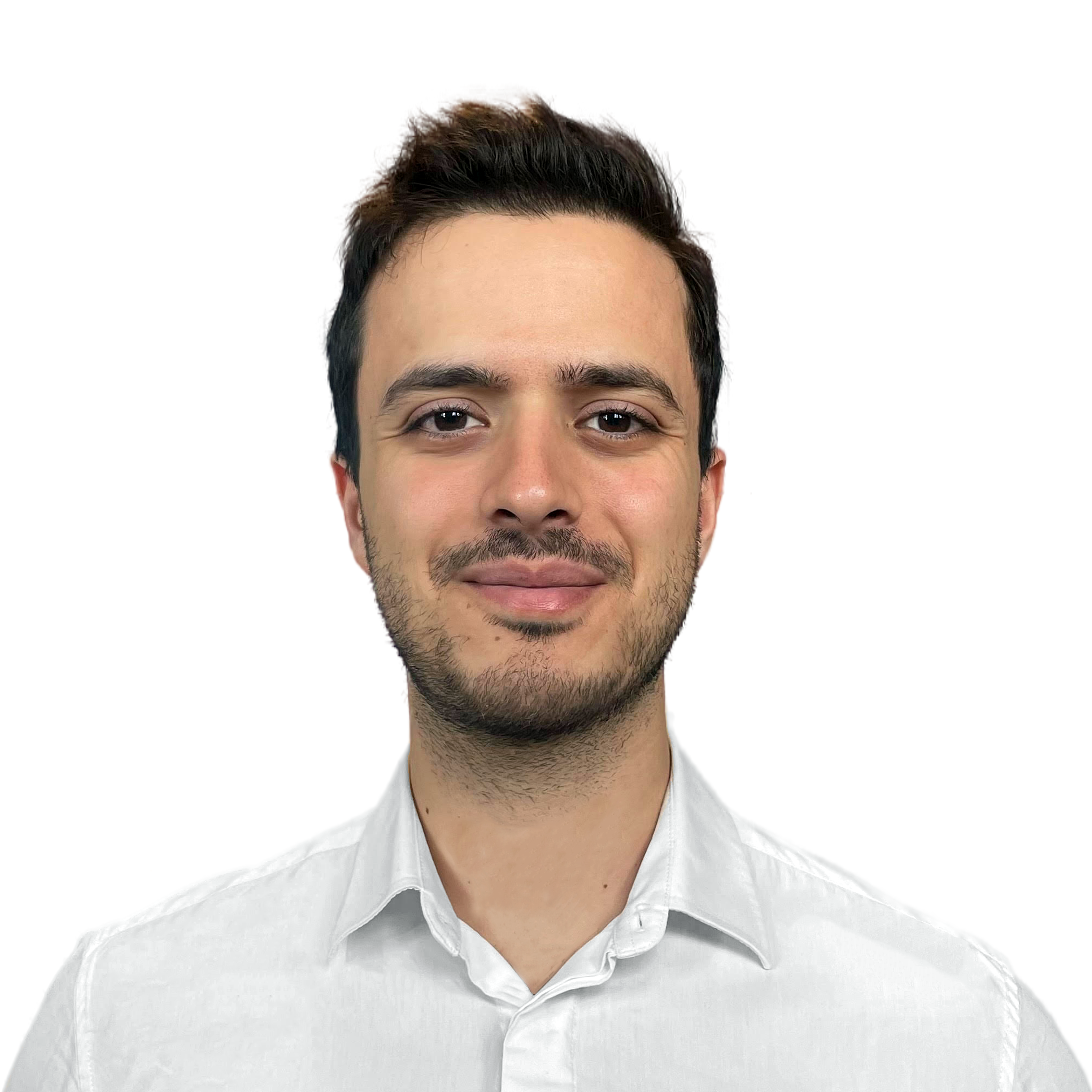}}]{Michael Khayyat} received the B.Eng degree in mechanical engineering from the American University of Beirut, Lebanon in 2016 and the M.Sc degree in mechanical engineering from Politecnico di Milano, Italy in 2019. His research interest lie in the field of game-theoretic motion planning and optimal control. He is currently a doctoral candidate at Politecnico di Milano.
\end{IEEEbiography}
\begin{IEEEbiography}[{\includegraphics[width=1in,height=1.25in,clip,keepaspectratio]{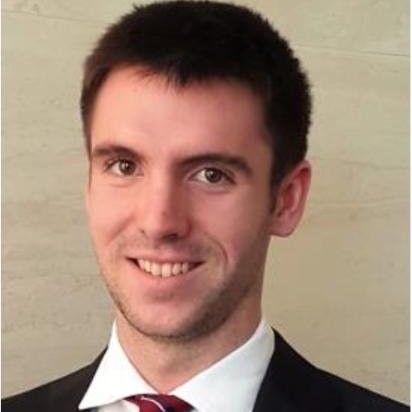}}]{Alessandro Zanardi} received 
the B.Sc. degree in automation and control engineering from Politecnico di Milano, M.Sc. degree in ``Robotics, Systems, and Control'' from ETH Zurich, and a Ph.D. degree in Mechanical Engineering (Robotics) from ETH Zurich in 2015, 2017 and 2023, respectively. His research interests include the development and evaluation of decision-making processes for embodied AI in multi-agent environments. This entails developing game-theoretic motion planning algorithms, modeling complex risks and uncertainties, and assessing behaviors in the context of safety-critical decision-making. He is currently a postdoctoral researcher at ETH Zurich. 
\end{IEEEbiography}
\begin{IEEEbiography}[{\includegraphics[width=1in,height=1.25in,clip,keepaspectratio]{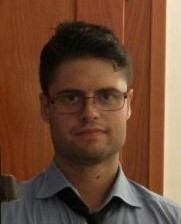}}]{Stefano Arrigoni}
received the M.S. degree in mechanical engineering and the Ph.D. degree in applied mechanics from Politecnico di Milano, Milano, Italy, in 2013 and 2017, respectively. His research interests lie in the area of autonomous vehicles with a focus on motion planning techniques and V2V communication. He is currently an assistant professor at the Department of Mechanical Engineering at Politecnico di Milano.
\end{IEEEbiography}
\begin{IEEEbiography}[{\includegraphics[width=1in,height=1.25in,clip,keepaspectratio]{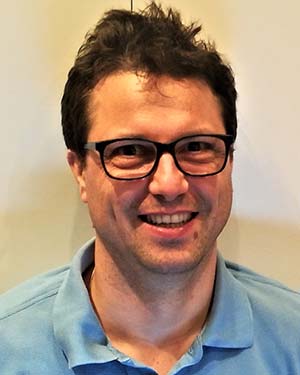}}]{Francesco Braghin}
(Member, IEEE) received the M.Sc. degree in mechanical engineering in 1997, and the Ph.D. degree in applied mechanics in 2001 from the Politecnico di Milano, Milan, Italy, where he has been a Full Professor since 2015. He is the Author of more than 250 scientific publications, his research is carried out in the field of vehicle dynamics and mechatronics. In particular, as regards road vehicles, his research deals with the modelling of tires and their interaction with the soil, and the application of optimal control algorithms to the design of hybrid and electric vehicles, and the development of fully autonomous vehicles.
\end{IEEEbiography}

\clearpage 

\appendix

\setcounter{table}{0}
\makeatletter 
\renewcommand{\thetable}{S\@Roman\c@table}
\makeatother

\setcounter{figure}{0}
\makeatletter 
\renewcommand{\thefigure}{S\@arabic\c@figure}
\makeatother

\setcounter{algorithm}{0}
\makeatletter 
\renewcommand{\thealgorithm}{S\@arabic\c@algorithm}
\makeatother

\renewcommand{\theHtable}{Supplement.\thetable}
\renewcommand{\theHfigure}{Supplement.\thefigure}

\subsection{Homotopy Diagram Illustrations}
\begin{figure}[H]
    \centering
    \includegraphics[scale=0.85]{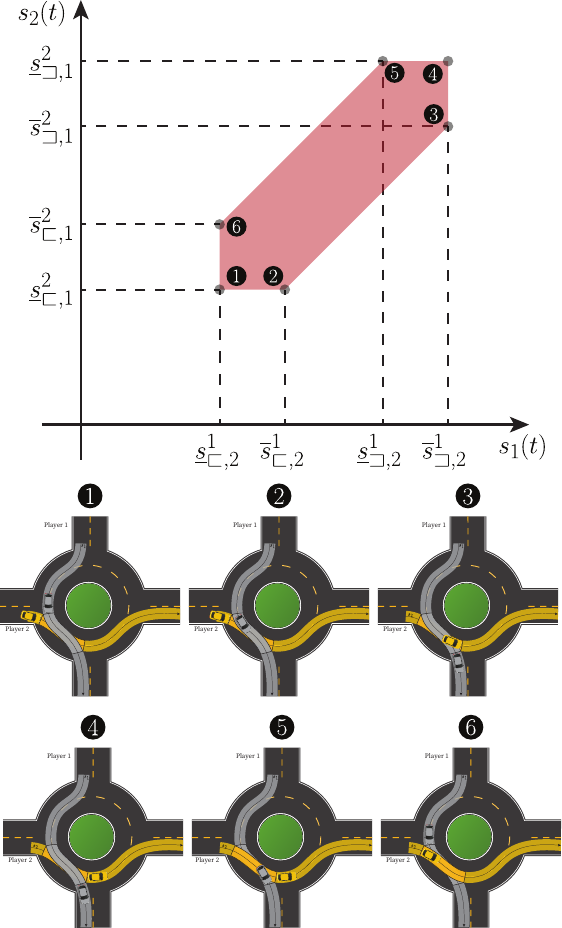}
    \caption{Different points labeled (1-6) in the collision area progress plane ($s_1, s_2$) and the corresponding positions of vehicles in the global frame with respect to their respective conflict region.}
    \label{fig:additionalhomotopy}
\end{figure}
\vfill

\subsection{Refined Formulation to Collision Avoidance Constraints}
\label{appendix:scenario_b}

Collision avoidance between \textit{combinations} of any two players $\nu, \mu$ whose path envelopes intersect can be formulated as a Mixed-Integer Problem (MIP). We introduce three binary variables, denoted as $\sigma^{\nu,\mu}_{A,F}$, $\sigma^{\nu,\mu}_{B,E}$ and $\sigma^{\nu,\mu}_{C,D}$ which take the following values:

\begin{equation*}
    \sigma^{\nu,\mu}_{\square,\triangle}(k)  = \begin{cases}
        1 & \text{if \myCircled{$\square$} or \myCircled{$\triangle$} are active} \\
        0 & \text{otherwise.} \\        
    \end{cases}
\end{equation*}

Utilizing the Big-$M$ formulation, the collision-avoidance constraints can be formulated as follows between two players $\nu$ and $\mu$:

\begin{subequations}
\label{eq:appendixmiqcp}
\begin{align}
&\begin{aligned}s^\mu (k) &\leq \underline{s}^{\mu}_{\sqsubset,\nu} +  M \cdot (1- h^{\nu,\mu})\cdot (1 - \sigma^{\nu,\mu}_{A,F}(k) ) \\  
& \quad + M \cdot h^{\nu,\mu}
\end{aligned} \label{eqn:collision2_a} \\
&\begin{aligned} s^\mu (k) &\leq s^{\nu}(k) + (\underline{s}^{\mu}_{\sqsubset,\nu} - \overline{s}^{\nu}_{\sqsubset,\mu} ) + M \cdot h^{\nu,\mu}(k) \\
& \quad+ M \cdot (1- h^{\nu,\mu})\cdot (1- \sigma^{\nu,\mu}_{B,E}(k))  \end{aligned}\label{eqn:collision2_b} \\
&\begin{aligned}s^\nu (k) &\geq \overline{s}^{\nu}_{\sqsupset,\mu} - M \cdot (1- h^{\nu,\mu})\cdot (1 - \sigma^{\nu,\mu}_{C,D}(k)) \\
& \quad - M \cdot h^{\nu,\mu}  \end{aligned}\label{eqn:collision2_c} \\
&\begin{aligned}s^\nu (k) &\leq \underline{s}^{\nu}_{\sqsubset,\mu}  + M \cdot  h^{\nu,\mu} \cdot(1 - \sigma^{\nu,\mu}_{C,D}(k)) \\
& \quad+ M \cdot(1 - h^{\nu,\mu}) \end{aligned}\label{eqn:collision2_d} \\
&\begin{aligned} s^\nu (k) &\leq s^{\mu}(k) - (\overline{s}^{\mu}_{\sqsubset,\nu} - \underline{s}^{\nu}_{\sqsubset,\mu} )  + M \cdot(1 - h^{\nu,\mu}) \\
& \quad + M \cdot   h^{\nu,\mu} \cdot (1- \sigma^{\nu,\mu}_{B,E}(k))\end{aligned}\label{eqn:collision2_e} \\
&\begin{aligned}s^\mu (k) &\geq \overline{s}^{\mu}_{\sqsupset,\nu} -  M \cdot h^{\nu,\mu} \cdot (1 - \sigma^{\nu,\mu}_{A,F}(k) ) \\
& \quad - M \cdot(1 - h^{\nu,\mu}) \end{aligned}\label{eqn:collision2_f} \\
&\sigma^{\nu,\mu}_{A,F}(k) + \sigma^{\nu,\mu}_{B,E}(k) + \sigma^{\nu,\mu}_{C,D}(k) = 1\label{eq:collision2_g}
\end{align}
\end{subequations}

(\ref{eq:collision2_g}) ensures that at least one of the constraints is active. To further reduce the search space, we can add additional constraints. Since we assume that all players do not drive back, then it is straightforward to see that for $h^{\nu,\mu}=0$, $\sigma^{\nu,\mu}_{A,F}$ is non-increasing, and $\sigma^{\nu,\mu}_{C,D}$ is non-decreasing. Meanwhile, for $h^{\nu,\mu}=1$, $\sigma^{\nu,\mu}_{A,F}$ is non-decreasing, and $\sigma^{\nu,\mu}_{C,D}$ is non-increasing. This cannot be said about $\sigma^{\nu,\mu}_{B,E}$, as it exhibits non-monotonic behavior. This can be introduced in the formulation through the use of slack variables $\xi^{\nu,\mu}_{A,F}$ and $\xi^{\nu,\mu}_{C,D}$, as follows:

\begin{equation}
    \begin{split}
        \sigma^{\nu,\mu}_{A,F}(k) - \sigma^{\nu,\mu}_{A,F}(k+1) - \xi^{\nu,\mu}_{A,F}(k+1) &= 0 \\        
        \sigma^{\nu,\mu}_{C,D}(k+1) - \sigma^{\nu,\mu}_{C,D}(k) - \xi^{\nu,\mu}_{C,D}(k+1) &= 0 \\
        \xi^{\nu,\mu}_{A,F}(k) - M\cdot h^{\nu,\mu} &\leq 1 \\
        -\xi^{\nu,\mu}_{A,F}(k) - M\cdot h^{\nu,\mu} &\leq 0 \\
        \xi^{\nu,\mu}_{A,F}(k) + M\cdot h^{\nu,\mu} &\leq M \\
        -\xi^{\nu,\mu}_{A,F}(k) - M\cdot h^{\nu,\mu} &\leq 1 +M \\
        \xi^{\nu,\mu}_{C,D}(k) - M\cdot h^{\nu,\mu} &\leq 1 \\
        -\xi^{\nu,\mu}_{C,D}(k) - M\cdot h^{\nu,\mu} &\leq 0 \\
        \xi^{\nu,\mu}_{C,D}(k) + M\cdot h^{\nu,\mu} &\leq M \\
        -\xi^{\nu,\mu}_{C,D}(k) + M\cdot h^{\nu,\mu} &\leq 1 +M 
    \end{split}
\end{equation}

such that $\xi^{\nu,\mu}_{A,F}, \xi^{\nu,\mu}_{C,D} \in [-1,1]$

Note that (\ref{eq:appendixmiqcp}) results in a Mixed-Integer Quadratically Constrained Problem (MIQCP), as a result of the multiplication of the binary variables.  Nonetheless, we can convert it into a MIQP using the translation of the if-then-else logic relation from~\cite{bemporad2001discrete}. To this end, we introduce three continuous auxiliary variables $\epsilon^{\nu,\mu}_{A,F}$, $\epsilon^{\nu,\mu}_{B,E}$ and $\epsilon^{\nu,\mu}_{C,D}$, such that (\ref{eq:appendixmiqcp}) is transformed into the following equivalent form:

\begin{subequations}
\label{eq:appendixmiqcpintomiqp}
\begin{align}
&s^\mu (k) \leq \underline{s}^{\mu}_{\sqsubset,\nu} +  M\cdot(1  - \sigma^{\nu,\mu}_{A,F}(k))  + M\cdot\epsilon^{\nu,\mu}_{A,F}(k) \label{eqn:collision2_a} \\
&\begin{aligned}s^\mu (k) &\leq s^{\nu}(k) + (\underline{s}^{\mu}_{\sqsubset,\nu} - \overline{s}^{\nu}_{\sqsubset,\mu} ) + M\cdot(1- \sigma^{\nu,\mu}_{B,E}(k))  \\
&\quad+ M\cdot \epsilon^{\nu,\mu}_{B,E}(k)\end{aligned}\label{eqn:collision2_b} \\
& s^\nu (k) \geq \overline{s}^{\nu}_{\sqsupset,\mu} - M \cdot(1 - \sigma^{\nu,\mu}_{C,D}(k)) + M \cdot \epsilon^{\nu,\mu}_{C,D}(k)\label{eqn:collision2_c} \\
&s^\nu (k) \leq \underline{s}^{\nu}_{\sqsubset,\mu}  + M \cdot (1 - \epsilon^{\nu,\mu}_{C,D}(k)) \label{eqn:collision2_d} \\
&s^\nu (k) \leq s^{\mu}(k) - (\overline{s}^{\mu}_{\sqsubset,\nu} - \underline{s}^{\nu}_{\sqsubset,\mu} )  + M \cdot (1- \epsilon^{\nu,\mu}_{B,E}(k))\label{eqn:collision2_e} \\
&s^\mu (k) \geq \overline{s}^{\mu}_{\sqsupset,\nu} -  M \cdot (1 - \epsilon^{\nu,\mu}_{A,F}(k) )\label{eqn:collision2_f} \\
&\sigma^{\nu,\mu}_{A,F}(k) + \sigma^{\nu,\mu}_{B,E}(k) + \sigma^{\nu,\mu}_{C,D}(k) = 1\label{eq:collision2_g} \\
& \epsilon^{\nu,\mu}_{A,F}(k) \leq h^{\nu,\mu} \\
&  \epsilon^{\nu,\mu}_{A,F}(k) \leq \sigma^{\nu,\mu}_{A,F}(k) \\
&  \epsilon^{\nu,\mu}_{A,F}(k) \geq  \sigma^{\nu,\mu}_{A,F}(k) - (1 -  h^{\nu,\mu})\\
&  \epsilon^{\nu,\mu}_{B,E}(k) \leq h^{\nu,\mu} \\
& \epsilon^{\nu,\mu}_{B,E}(k) \leq \sigma^{\nu,\mu}_{B,E}(k) \\
& \epsilon^{\nu,\mu}_{B,E}(k)  \geq \sigma^{\nu,\mu}_{B,E}(k) - (1 -  h^{\nu,\mu})\\
& \epsilon^{\nu,\mu}_{C,D}(k) \leq h^{\nu,\mu} \\
&  \epsilon^{\nu,\mu}_{C,D}(k) \leq \sigma^{\nu,\mu}_{C,D}(k) \\
&  \epsilon^{\nu,\mu}_{C,D}(k) \geq \sigma^{\nu,\mu}_{C,D}(k) - (1 - h^{\nu,\mu})
\end{align}
\end{subequations}

where $\epsilon^{\nu,\mu}_{\square,\triangle}$ is effectively equal to $h^{\nu,\mu}\cdot\sigma^{\nu,\mu}_{\square,\triangle}$

\subsection{Identification of Deadlock Cases}
\label{deadlock}

Not all scenario homotopy classes, i.e passing sequences, are physically feasible due to deadlocks. This kind of \textit{deadlock} arises when the path envelopes of three or more vehicles intersect each other, and occurs purely due to the geometry of the paths and the the monotonicity constraint on the progress of players (\ref{eq:stateandcontrolconstraints_b}). Depending on the application of the presented framework, pre-identification of deadlock scenario homotopy classes could prove advantageous.

It is possible to identify deadlock cases by attempting to solve a constraint satisfaction problem (CSP) involving a relatively small number of points $N$, depending on the geometric complexities of the scenario. We have found that satisfactory outcomes are attainable with $N$ values as small as the number of components within the scenario homotopy class, as follows:

\begin{equation}
  \label{eq:csp}
\left\{
\begin{split}
&\underset{s^\nu(k)~\forall \nu \in \mathcal{A}~\forall k\in [0,\dots,N]}{\min} \    0    & \\[2ex]
&\text{s.t } \ \begin{array}[t]{ll} 
& s^\nu(k+1) \geq s^\nu(k) \ \forall k \in [0,\dots,N-1] ~ \forall \nu \in \mathcal{A} \\
& s^\nu(0) \leq \min([\underline{s}^\nu_{\sqsubset,\mu}]_{\forall {\mu \in \mathcal{I}^\nu}})~ \forall \nu \in \mathcal{A} \\
& s^\nu(N)  \geq \max([\bar{s}^\nu_{\sqsupset,\mu}]_{\forall {\mu \in \mathcal{I}^\nu}})~\forall \nu \in \mathcal{A} \\
& \langle [h^{\nu,\mu}]_{\forall \mu \in \mathcal{I}^\nu,~\forall \nu \in \mathcal{A}} \rangle\\
&(\ref{eq:collision}), (\ref{eq:logical}), (\ref{eq:monotonicity}), (\ref{eq:collision2})
                                            \end{array} 
\end{split}
\right.
\end{equation}

\subsection{Demonstration of \cref{fig:homotopy_otherscenarios_d}}
\label{appendix:scenario_d}
\begin{figure}[H]
    \centering
    \includegraphics[scale=0.85]{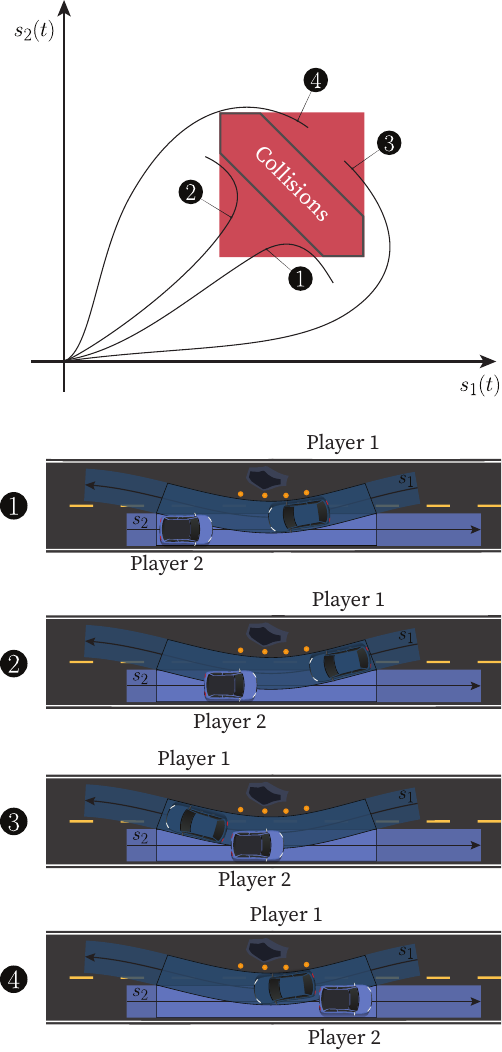}
    \caption{Illustrates the effective collisions area in the case where $\text{sgn}(s^\nu_{\otimes2,\mu} - s^\nu_{\otimes1,\mu}) \neq \text{sgn}(s^\mu_{\otimes2,\nu} - s^\mu_{\otimes1,\nu})$ that arises due to the monotonicity constraint on the progress of the each player. Note that in cases (1-4), the progress of either player would be impossible unless the other player drives in reverse ($\dot{s}^\nu < 0$), violating the monotonicity constraint.}
    \label{fig:appendix_fig_constraints}
\end{figure}

\end{document}